# Observation of half-height magnetization steps in $Sr_2RuO_4$

J. Jang[1], D.G. Ferguson[1], V. Vakaryuk[1,2], R. Budakian[1*], S.B. Chung[3], P.M. Goldbart[1], and Y. Maeno[4]

1. Department of Physics, University of Illinois at Urbana-Champaign, Urbana, IL 61801-3080

2. Materials Science Division, Argonne National Laboratory, Argonne IL 60439

3. Department of Physics, McCullough Building, Stanford University, Stanford, CA 94305-4045

4. Department of Physics, Kyoto University, Kyoto 606-8502, Japan

* To whom correspondence should be addressed. E-mail: budakian@illinois.edu

Spin-triplet superfluids can support exotic objects, such as half-quantum vortices characterized by the nontrivial winding of the spin structure. We present cantilever magnetometry measurements performed on mesoscopic samples of $Sr_2RuO_4$, a spin-triplet superconductor. For micron-sized annular-shaped samples, we observe transitions between integer fluxoid states, as well as a regime characterized by "half-integer transitions," i.e., steps in the magnetization with half the height of the ones we observe between integer fluxoid states. These half-height steps are consistent with the existence of half-quantum vortices in superconducting $Sr_2RuO_4$.

Most known superconductors are characterized by the spin-singlet pairing of the electrons that constitute the superconducting flow. An exception is $Sr_2RuO_4$ (SRO) which, much like the A-phase of superfluid $^3$He, may exist in the equal-spin pairing (ESP) phase (*1*). This phase has been proposed to host half-quantum vortices (HQVs), which are characterized by the relative winding of the phase of the spin-up and spin-down components of the superfluid order parameter (*2-3*). In addition to being of basic scientific interest, HQVs are expected to give rise to zero-energy Majorana quasiparticles (*4-5*), which have been suggested as a resource for topological quantum computation (*6*).

The ESP state may be thought of as comprising two weakly interacting condensates, having Cooper-pair spin configurations, $|\uparrow\uparrow\rangle$ and $|\downarrow\downarrow\rangle$, defined with respect to a common (i.e., ESP) axis. An HQV corresponds to the winding of the phase of only one of these condensates around a contour that encircles the HQV core, e.g., $(\Delta\theta_\uparrow, \Delta\theta_\downarrow) = (\pm 2\pi, 0)$ or $(0, \pm 2\pi)$, producing half of the magnetic moment of a conventional (i.e., full-quantum) vortex (FQV), for which $\Delta\theta_\uparrow = \Delta\theta_\downarrow = \pm 2\pi$. The Meissner response of the superconductor screens charge currents over the lengthscale of the London penetration depth λ; however, any (charge-neutral) spin currents go unscreened. Consequently, the kinetic energy of an isolated HQV diverges logarithmically with the system size, whereas the kinetic energy of an FQV would remain finite. Hence, a single HQV may not be energetically stable in a macroscopic sample, whereas, according to (*7*), a single HQV could be stable in a mesoscopic SRO sample of size comparable to or smaller than *λ*.

We use cantilever magnetometry to measure the magnetic moment of micron-sized SRO samples, with the aim of distinguishing between HQV and FQV states via changes in magnetic moment associated with the entry of single vortices. To facilitate this aim, we have fabricated annular samples by drilling a hole in the center of each particle using a focused ion beam. This geometry yields a discrete family of equilibrium states, in which the order parameter winds around the annulus as it would around a vortex core, but evades complications arising from the vortex core.

For an annular conventional superconductor, the fluxoid $\Phi'$, defined via $\Phi' = \Phi + (4\pi/c)\oint \lambda^2 \vec{j}_s \cdot d\vec{s} = n\Phi_0$, must be an integer multiple $n$ of the flux quantum $\Phi_0 = hc/2e$ for any path encircling the hole (8). Here, $\vec{j}_s$ is the supercurrent density, $\Phi = \oint \vec{A} \cdot d\vec{s}$ is the magnetic flux enclosed by the path, $\vec{A}$ is the vector potential, and $n = \oint \vec{\nabla}\theta \cdot d\vec{s}/2\pi$ is the order-parameter winding-number along the path. In the regime where the wall thickness of the annulus becomes comparable to or smaller than $\lambda$, $\vec{j}_s$ will not necessarily vanish in the interior of the annulus; hence, it is the fluxoid and not the flux that is quantized. The quantized winding of the order parameter, however, produces observable effects in the magnetic response of the annulus. In the regime in which the magnetization is piecewise linear in the magnetic field, the supercurrents that flow around the hole produce a magnetic moment $\mu_z = \Delta\mu_z n + \chi_M H_z$, where $\chi_M$ is the Meissner susceptibility and $H_z$ is the component of the magnetic field that controls the flux through the hole (Fig.1). In equilibrium, changes in the fluxoid are associated with transitions in the winding number in single units (i.e., $n \to n+1$), corresponding to the changes in the magnetic moment in increments of $\Delta\mu_z$.

For an ESP superconductor, the two condensates bring the two integer-valued winding numbers: $n_\uparrow$ and $n_\downarrow$. Then, the role of $n$ is played by the half-sum $n = (n_\uparrow + n_\downarrow)/2$. The integer-fluxoid (IF) state of the annulus—the coreless analog of the FQV state—corresponds to the common winding of the condensates (i.e., $n_\uparrow = n_\downarrow$), whereas the half-fluxoid (HF) state—the coreless analog of the HQV state—corresponds to winding numbers that differ by unity (i.e., $n_\uparrow = n_\downarrow \pm 1$). Thus, equilibrium transitions between the IF and HF states would change $n$ by half a unit (i.e., $n \to n \pm 1/2$), and this would produce a change $\Delta\mu_z/2$ of the magnetic moment, i.e., half of that produced for an equilibrium transition between two IF states (*9*).

To measure the magnetic response of the superconductor, we use a recently developed phase-locked cantilever magnetometry technique (*10*) operating inside a $^3$He refrigerator with a base temperature of 300 mK (Fig. 1A). In our setup, the displacement of the cantilever is measured using a fiber optic interferometer operating at a wavelength of 1510 nm. The optical power is maintained at 5 nW in order to minimize heating of the sample due to optical absorption. To exclude the potential fortuitous effects of geometry, we present data for three annular SRO particles, of different sizes, fabricated from a high-quality SRO single crystal with a bulk transition temperature $T_c$ = 1.43 K, grown using the floating-zone method (*11*).

We start with the particle shown in Fig. 1A. To characterize its equilibrium fluxoid state, the particle is heated above $T_c$ by momentarily increasing the laser power, and then cooled below $T_c$ in the presence of a static magnetic field (i.e., field cooling). The field-cooled data (Fig. 1C) exhibits periodic steps in the magnetic moment, of nearly constant magnitude $\Delta\mu_z = (4.4 \pm 0.1)\times 10^{-14}$ e.m.u., period $\Delta H_z = (16.1 \pm 0.1)$ Oe, and susceptibility $\chi_M = -6.0\times 10^{-15}$ cm$^3$. The measured period $\Delta H_z$ is in reasonable agreement with theoretical estimates for the equilibrium fluxoid transitions of a hollow superconducting cylinder having the dimensions of the SRO sample (estimated value $\Delta H_z \approx 20$ Oe) (*12*). Thus, we conclude that the periodic events observed in Fig. 1C correspond to equilibrium transitions between IF states of the annular SRO particle.

The presence of an in-plane magnetic field $H_x$ brings two new features: (i) For $H_z = 0$, the in-plane magnetic response of the sample exhibits a Meissner behavior for $H_x < 250$ Oe. At $H_x = 250$ Oe, we observe a step in the in-plane magnetic moment with magnitude $\Delta\mu_x \approx 2 \times 10^{-14}$ e.m.u.; both the magnitude of the step and the value of $H_x$ at which it occurs are consistent with those expected for the critical field $H_{c1} \parallel ab$ and $\Delta\mu_{ab}$ of an in-plane vortex for our micron-size sample (Fig. S3). (ii) For $H_x < H_{c1} \parallel ab$, where no in-plane vortices are expected to penetrate the sample, and in the presence of $H_z$, we observe the appearance of half-integer (HI) states, for which the change in the magnetic moment of the particle is half that of the IF states.

Figure 2A shows data taken at $T = 0.6$ K for in-plane fields $H_x$ ranging from -200 to 200 Oe. The data in Fig. 2A were obtained by cooling the sample through $T_c$ in zero field and performing a cyclic field sweep starting at $H_z = 0$ (i.e., zero-field cooling). At this temperature, the zero-field cooled and field cooled data are nearly identical, indicating that the equilibrium response is well-described by the zero-field cooled data. Figure 2B shows the data after subtracting the Meissner response; Fig. 2C shows a histogram of the data in Fig. 2B. The steps, which remain, indicate that the change in magnetic moment associated with HI → IF transitions is half of that associated with transitions between two IF states (Table 1). Figure S4 shows zero-field cooled data taken at $T = 0.5$ K for a direction of the in-plane field having been rotated by 35° in the $ab$-plane of the sample. We find that the half-step features remain and, furthermore, that the range of $H_z$ values for which the HI state is the equilibrium state is influenced by the magnitude but not the direction of the in-plane field.

To verify that the HI features we observe correspond to fluxoid states, and not tilted or kinked vortex lines that pierce the sample walls, we performed an additional series of measurements on a particular SRO annulus to determine the dependence of the HI state on the sample geometry. Prior to each of these measurements, we cut away more of the annulus, using the focused ion beam. The motivation for this study was the following: the location and stability of a vortex line passing through the bulk of the sample should be sensitive to the sample geometry (e.g., the thickness of the walls of the annulus, or the location of pinning sites). By contrast, if the currents responsible for the half-step features are generated by a half-integer fluxoid, and thus only circulate the hole, the observed fractions should not be affected by the sample dimensions. Figure S7 shows measurements on a second SRO annulus; we again find that the in-plane magnetic field stabilizes an HI state in which the observed fraction is very nearly a half ($0.50 \pm 0.02$). Figure S8B shows the image of the sample presented in Fig. S7 (black outline), as well as the outline after reshaping (purple outline); after reshaping, the sample volume was reduced to 44% of the original volume, however the half-integer fraction was not affected ($0.50 \pm 0.01$). We find that half-height step features observed in these samples are robust, and not sensitive to the wall thickness or the shape of the boundary.

We have also studied annular samples for which the HI state is not expected to occur: (a) an SRO particle whose dimensions are considerably larger than $\lambda$ (Fig. S9), and (b) a micron-size particle fabricated from $NbSe_2$—a spin-singlet, layered superconductor (Fig. S10). For both of these samples we find that, as the applied field is increased, a complicated set of fractional steps in the magnetic moment emerges—their fraction need not be one-half, and it changes with field. Furthermore, the pattern of fractional steps depends on the direction of the in-plane field (i.e., changes when $H_x \to -H_x$). The irregular pattern of fractional steps found for these particles is consistent with the presence of vortices in the bulk of the sample.

The temperature dependence of the fractional steps measured for the large SRO (Fig. 3A) and NbSe$_2$ (Fig. 3B) samples show qualitatively different behavior from that of the HI steps observed for the smaller SRO sample, shown in Fig. 1 (Fig. 3C). As the temperature is raised towards $T_c$, the fractional steps observed in Figs. 3A and 3B become less pronounced and, eventually, most disappear, leaving only the periodic fluxoid transitions. Numerical simulations for thin superconducting discs containing a circular hole (*13*) find that as $\lambda$ and the coherence length $\xi$ become comparable to or larger than the wall thickness of the ring, the fluxiod states become favored energetically over bulk vortices (i.e., vortices penetrating the walls of the superconductor). Thus, at higher temperatures, bulk vortices should be less stable, in part because, near $T_c$, $\xi$ and $\lambda$ will increase and eventually become large (relative to the wall thickness), and also because of increased thermal fluctuations. This behavior is consistent with the temperature dependence observed for the large SRO and NbSe$_2$ samples. In contrast, the HI transitions persist at higher temperature, and the relative contribution to the magnetic moment from each HI transition does not change significantly with temperature. Importantly, the HI transitions measured for the SRO sample shown in Fig. 1 exhibit a qualitatively similar temperature dependence to the fluxoid transitions: near $T_c$, the HI transitions become reversible and broaden (Fig. 3C), indicating that $\xi$ is comparable to the wall thickness in a portion of the ring. The HI transitions are clearly identified by the two double peaks in the derivative signal.

The half-integer states observed in magnetometry measurements performed on mesoscopic rings of SRO are consistent with the existence of half-quantum fluxoid states in this system. Our key findings—the reproducibility of the half-height steps in the *c*-axis magnetic moment in multiple samples, and their evolution with the applied magnetic field—demonstrate that the half-integer states are intrinsic to the small SRO annuli. These findings can be understood qualitatively on the basis of existing theoretical models of HQVs (SOM text). In addition to the magnetic response, further studies will probe characteristics that are particular to the HQV state, such as spin currents or vortices obeying non-abelian statistics (*14-15*).

16. We thank D. Van Harlingen, M. Stone, E. Fradkin, E.-A. Kim, and H. Bluhm for valuable discussions, and M. Ueda for helpful suggestions regarding the data analysis. In particular, the authors thank A.J. Leggett for his theoretical guidance. This work was supported by the U.S. Department of Energy Office of Basic Sciences, grant DEFG02-07ER46453 through the Frederick Seitz Materials Research Laboratory at the University of Illinois at Urbana Champaign and the Grant-in-Aid for the Global COE Program from MEXT of Japan.

**Figure legends**

**Figure 1:** Image of cantilever with attached annular SRO particle. (**A**) The 80 μm × 3 μm × 100 nm single-crystal silicon cantilever has natural frequency $\omega_0/2\pi = 16$ kHz, spring constant $k = 3.6 \times 10^{-4}$ N/m, and quality factor $Q = 65,000$, and exhibits a thermal-limited force sensitivity of $S_F^{1/2} \approx 1.0 \times 10^{-18}$ N/$\sqrt{Hz}$ at $T = 0.5$ K. (Inset) SEM of the 1.5 μm × 1.8 μm × 0.35 μm annular SRO sample attached to the cantilever. The orientation of the *ab* planes is clearly visible from the layering observed near the edges of the SRO particle. (**B**) Anisotropic component of the susceptibility $\Delta\chi = \chi_c - \chi_{ab}$ as a function of temperature. Here, $\chi_c$ and $\chi_{ab}$ are the *c*-axis and in-plane susceptibilities, respectively. (**C**) Field-cooled data measured at $T = 0.45$ K for $H_x = 0$.

**Figure 2:** Evolution of the half-integer state with in-plane magnetic field. (**A**) Zero-field-cooled data obtained at $T = 0.6$ K. (**B**) Data shown in (**A**) after subtracting the linear Meissner response; curves have been offset for clarity. (**C**) Histogram of the Meissner-subtracted data. The red points show the mean value of each cluster in the histogram, corresponding to the mean value of a given plateau; the horizontal error bars represent the standard deviation of a given cluster. The change in moment corresponding to the $i^{th}$ transition is labeled $\Delta_i$.

| $H_x$ (Oe) | $\Delta_1/(\Delta_1+\Delta_2)$ | $\Delta_3/(\Delta_3+\Delta_4)$ |
|---|---|---|
| 200 | 0.46 ± 0.06 | 0.46 ± 0.06 |
| 140 | 0.47 ± 0.05 | 0.47 ± 0.05 |
| 80 | 0.48 ± 0.06 | 0.51 ± 0.04 |
| -80 | 0.45 ± 0.06 | 0.53 ± 0.09 |
| -140 | 0.52 ± 0.07 | 0.51 ± 0.06 |
| -200 | 0.46 ± 0.07 | 0.43 ± 0.07 |
| | $\langle\Delta_1/(\Delta_1+\Delta_2)\rangle$ | $\langle\Delta_3/(\Delta_3+\Delta_4)\rangle$ |
| | 0.47 ± 0.03 | 0.48 ± 0.03 |

**Table 1:** The calculated fractional step heights for the data shown in Fig. 2C. The average value of a given fractional step is indicated by the quantity $\langle...\rangle$.

**Figure 3:** Temperature evolution of the fractional and HI states. The data were acquired at the value of the in-plane field indicated on the upper right-hand corner of each panel; all data were measured by field-cooling the samples. The Meissner response has been subtracted from all data; curves have been offset for clarity. (**A**) Data obtained for the NbSe$_2$ sample (Fig. S10). The data are scaled by: 4.5 K: 1.0×, 6.0 K: 1.7×, and 7.0 K: 10×. (**B**) Data obtained for the large SRO sample (Fig. S9). The data are scaled by: 0.63 K: 1.0×, 0.72 K: 1.5×, 0.76 K: 2.3×, and 0.80 K: 6.0×. (**C**) Data obtained for the SRO sample in Fig. 1. The $T = 0.55$ K data is a plot of the $c$-axis moment acquired by applying the phase-locked modulation of $\delta H_x = 1.0$ Oe perpendicular to the $c$-axis. For the $T \geq 0.80$ K data, we measure $d\mu_z/dH_z$ by applying the phase-locked modulation ($\delta H_z = 0.25$ Oe) parallel to the $c$-axis (SOM). The magnetic moment curves are calculated by integrating the measured derivative signal.

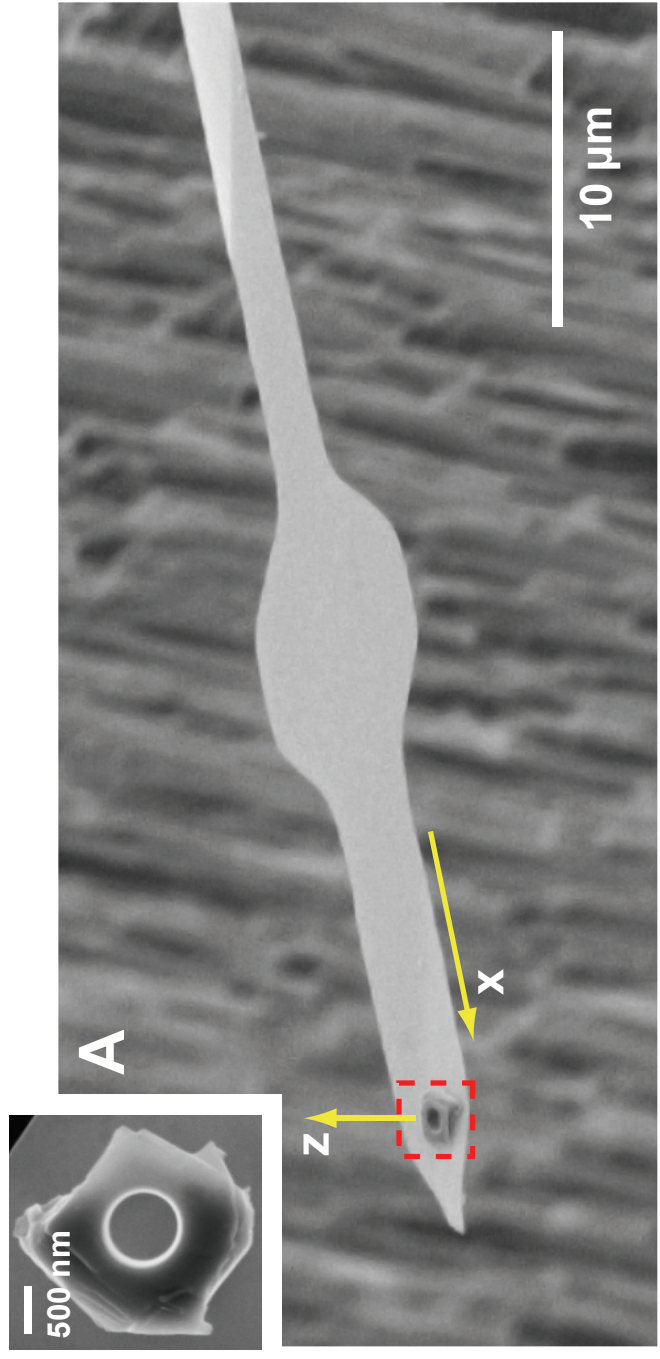
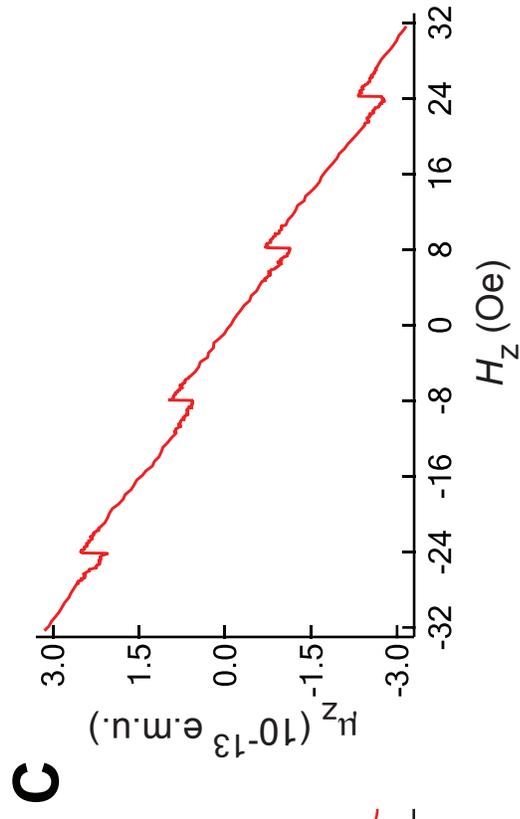
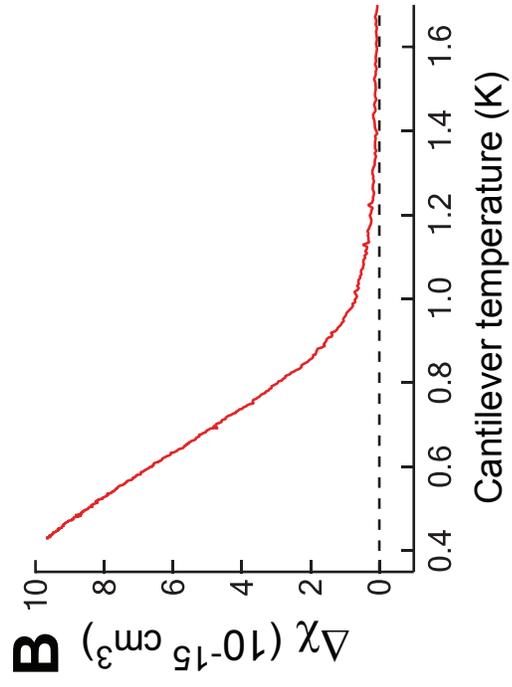

Fig. 1

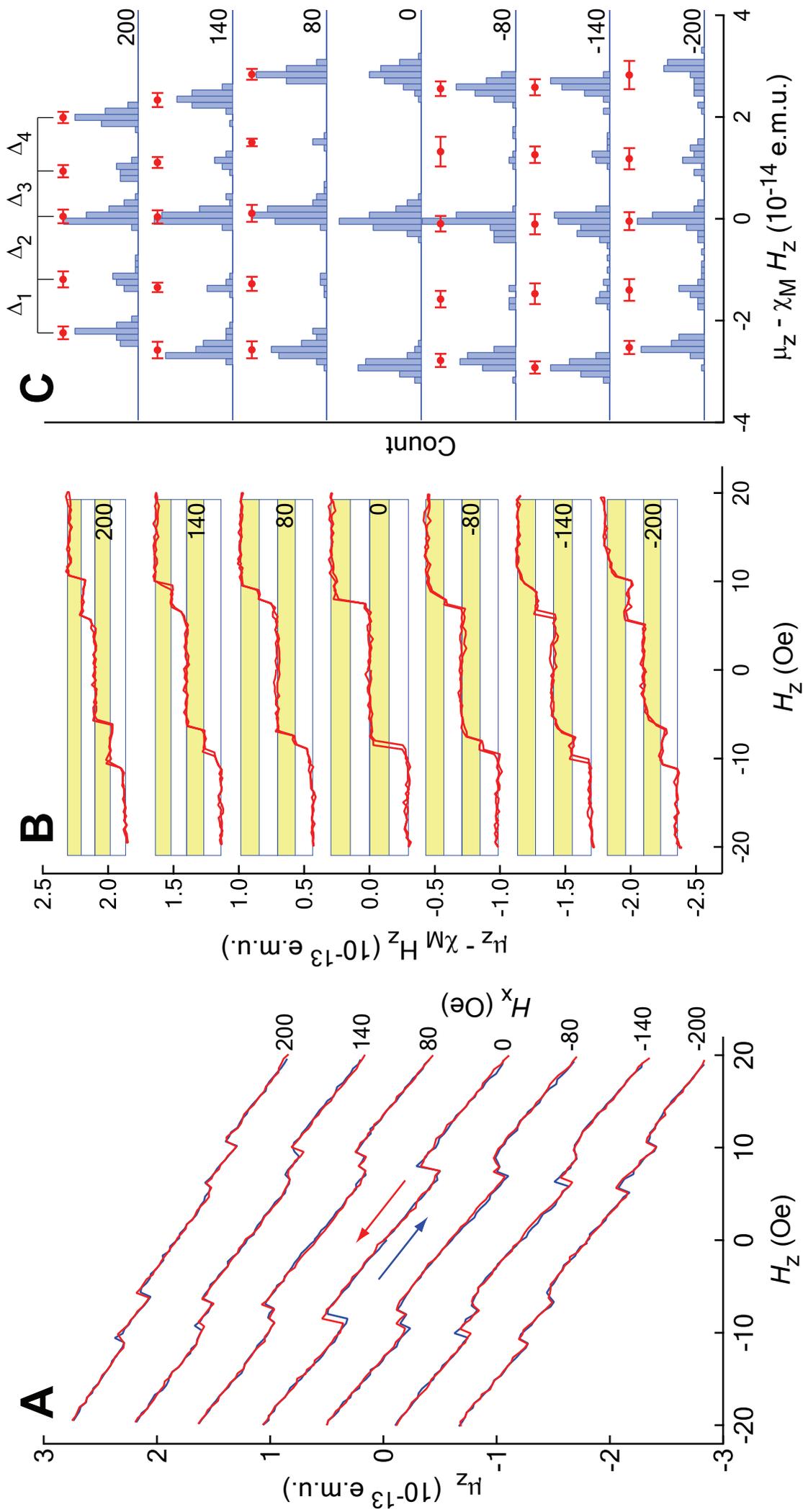

Fig. 2

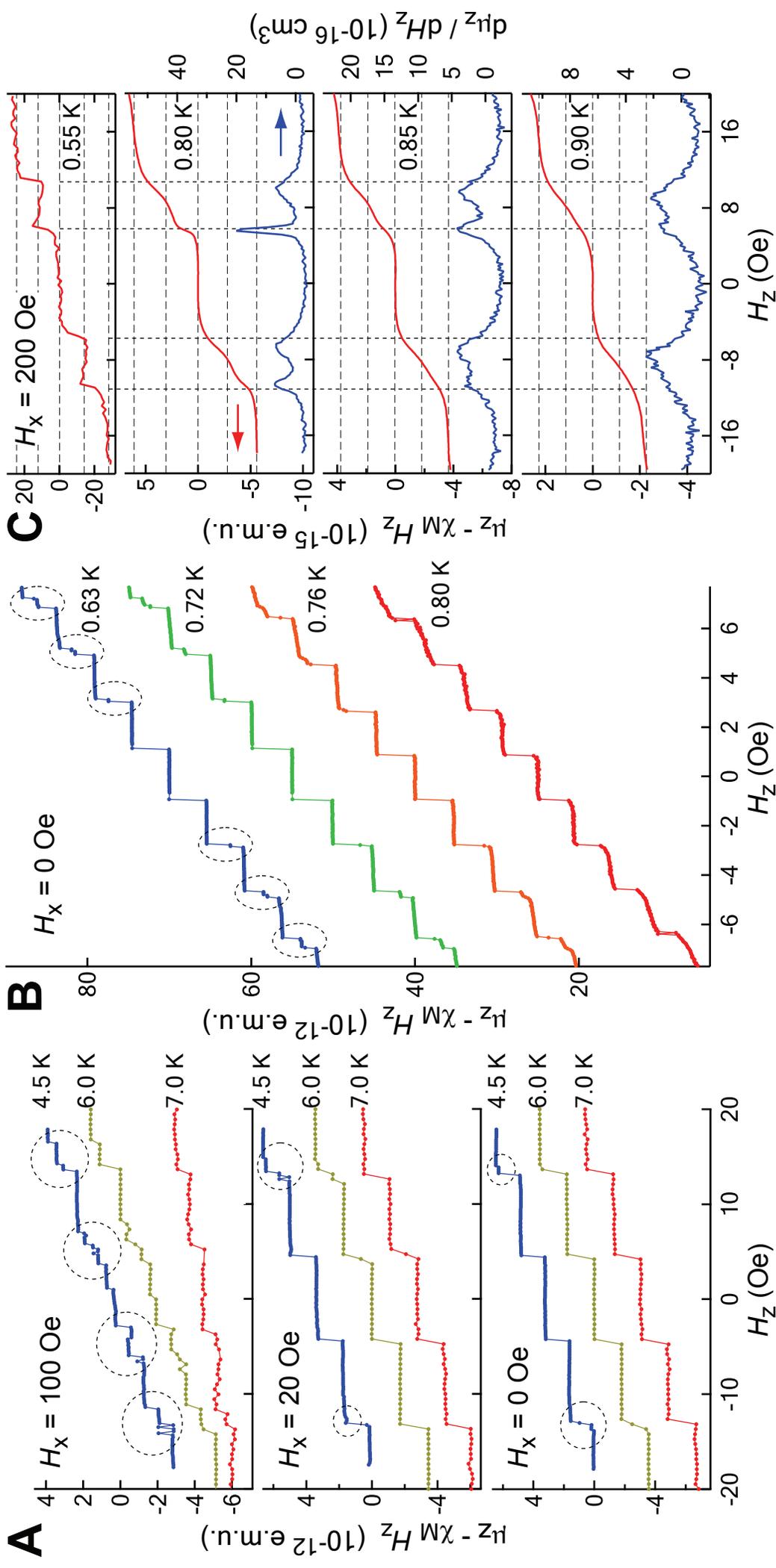

**Fig. 3**

Supporting Online Material for

# Observation of half-height magnetization steps in $Sr_2RuO_4$

J. Jang, D.G. Ferguson, V. Vakaryuk, R. Budakian[*], S.B. Chung, P.M. Goldbart, and Y. Maeno

* To whom correspondence should be addressed. E-mail: budakian@illinois.edu

**This PDF file includes:**

Materials and Methods

SOM Text

Figs. S1 to S10

Tables S1 and S2

References

## Materials and Methods

The micron-sized annular $Sr_2RuO_4$ samples were obtained from millimeter-sized crystals grown using the floating-zone method (*S1*). The SRO crystal was first glued to a post, then ~100-µm thick segments were cleaved using a razor blade. To avoid possible oxidation and surface contamination, only segments obtained from the interior of the larger crystal were used. The SRO segment was crushed on a silicon substrate and the pieces were imaged using the electron beam of a dual-column focused ion beam (FIB). The shape of the particles was found to be an important factor in determining the orientation of crystalline axis; for the particles used in this study, the orientation of the *ab* planes was clearly visible from the layering observed near the edges of the particle. Figure S1A is an example of a low magnification SEM image showing the particle distribution (after crushing), which was later used as a map to identify the location of the FIB-ed particle under optical microscope. Higher magnification SEM scans from various angles were necessary to confirm the layering of a particle. After locating the desired particle, the 30 kV ion source was used to cut a hole in the center of the particle parallel to the crystal *c*-axis. In order to minimize ion implantation, the number of images taken using the ion source was limited to two exposures at emission current of 1 pA; the hole was cut using an emission current of 10 pA. The milled particle was then transferred to a micro-manipulator stage of a long working distance Mitutoyo optical microscope. The FIB-ed particle was located from the pattern of the debris field imaged using the SEM. A three-axis hydraulic Narishige micromanipulator was used to position a pulled borosilicate micropipette having a < 1 μm tip diameter near the particle. The electrostatic interaction of the particle with the micropipette was sufficient to pick up the particle. The particle was glued to the tip of a custom-fabricated silicon cantilever. Prior to placement of the particle, a small amount of Gatan G-1 epoxy was placed on the tip of the cantilever. The epoxy was cured overnight at 70 °C in a nitrogen environment. The planar geometry of the particle ensured that the particle's *c*-axis was oriented perpendicular to the axis of the cantilever. However, the *a* and *b* axes had no special orientation with respect to the cantilever.

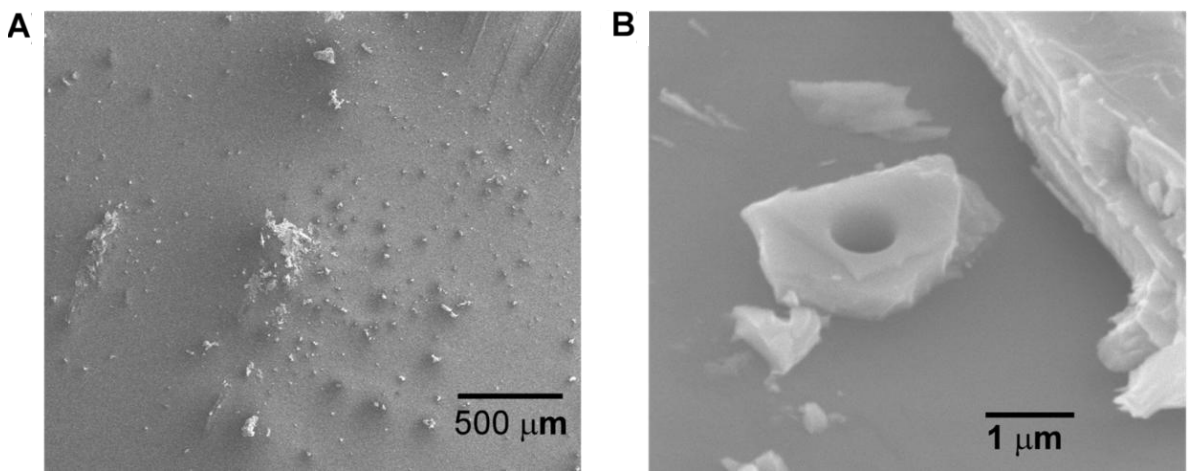

Figure S1: (**A**) Low magnification SEM of SRO debris field. (**B**) A micron-size SRO particle after cutting a hole using the FIB.

### Phase-locked cantilever magnetometry

In the presence of an external magnetic field $\boldsymbol{H}$, the cantilever is subject to a torque $\boldsymbol{\tau} = \boldsymbol{\mu} \times \boldsymbol{H}$ produced by the magnetic moment $\boldsymbol{\mu}(\boldsymbol{H})$ of the superconductor. During measurement, the cantilever is placed in a positive feedback loop, and is driven at its natural frequency using a piezoelectric transducer; the resulting tip motion is given by $z(t) = z_{pk} \cos \omega_c t$. To enhance detection sensitivity, a small time-dependent magnetic field $\delta H(t)$ is applied perpendicular to the component of $\boldsymbol{\mu}$ that we seek to detect. By making $\delta H(t)$ depend on the phase of the cantilever position, the cantilever experiences a dynamic position-dependent force, which shifts either the cantilever frequency or its dissipation. In general, the applied modulation can be of the form: $\delta \boldsymbol{H}(t) = \delta H_x \cos(\omega_c t + \theta_x)\hat{x} + \delta H_z \cos(\omega_c t + \theta_z)\hat{z}$, where the phase angles $\theta_x$ and $\theta_z$ can be independently chosen. In the presence of a static magnetic field $\boldsymbol{H} = H_x\hat{x} + H_z\hat{z}$ and the applied modulation $\delta \boldsymbol{H}(t)$, the Fourier transform of the equation of motion of the cantilever is given by:

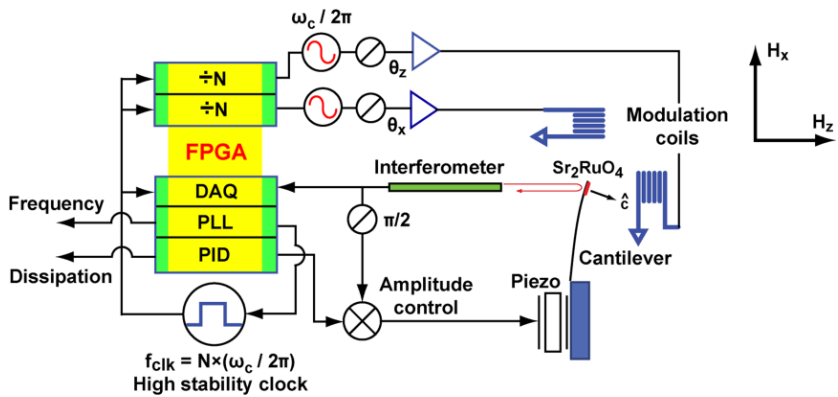

Figure S2: Schematic of phase-locked cantilever magnetometry apparatus.

$$(-m\omega^2 - im\gamma\omega + k)\tilde{z}(\omega) \tag{S1}$$

$$= \frac{1}{L_{eff}^2}\{(\mu_x - \chi_c H_x)H_x + (\mu_z - \chi_{ab}H_z)H_z\}\tilde{z}(\omega)$$

$$+ \frac{1}{L_{eff}z_{pk}}\{(\mu_x - \chi_c H_x)\delta H_z e^{i\theta_z} + (\mu_z - \chi_{ab}H_z)\delta H_x e^{i\theta_x}\}\tilde{z}(\omega).$$

Here, $k$, $m = k/\omega_c^2$, and $L_e (\approx L/1.38)$ are the cantilever spring constant, effective mass and effective length, respectively; $\mu_x$ and $\mu_z$ are the $c$-axis and $ab$-plane components of the magnetic moment and $\chi_c$ and $\chi_{ab}$ are the two components of the magnetic susceptibility of the sample. We note that for our samples $\chi_c \gg \chi_{ab}$. This assumption is justified because of the large shape anisotropy of our sample as well as the highly anisotropic response of the superconductor: $\gamma_{SC} = \lambda_c(0)/\lambda_{ab}(0) \approx 20$, where $\lambda_c(0) = 3.0 \times 10^3$ nm and $\lambda_{ab}(0) = 152$ nm are the zero temperature $c$-axis and $ab$-plane penetration depths, respectively (*S1*). In writing Eq. (S1), we have assumed that the peak angle of the cantilever deflection is $z_{pk}/L_{eff} \ll 1$. In our measurements, $L_{eff} \approx 55$ μm and $z_{pk} \approx 60$ nm, thus the assumption of small angles is justified. The static magnetic fields applied in the lab reference frame generate position-dependent fields in the oscillating reference frame of cantilever in addition to the applied modulation. To ensure that the magnetic field in the reference frame of the SRO sample has the desired time-dependence, both feedback and feed-forward modulation is applied in the x and z directions.

To measure the $c$-axis moment, we modulate the in-plane field by $\delta H_x = 1.0$ Oe; the relative phase is chosen so as to shift the cantilever dissipation $\gamma$: $\theta_x = \pi/2$. An expression for the shift in dissipation is obtained from Eq. (S1)

$$\Delta\gamma_z = \frac{\omega_c}{kL_{eff}z_{pk}}\delta H_x(\mu_z - \chi_{ab}H_z) \tag{S2}$$

In the regime that the superconductor exhibits linear response, $\mu_z = \Delta\mu_z n + \chi_c H_z$, where $n$ is the fluxoid quantum number and $\Delta\mu_z$ is the change in the magnetic moment of the ring associated with fluxoid entry. Since $\chi_c \gg \chi_{ab}$, Eq. (S2) reduces to

$$\Delta\gamma_z \approx \frac{\omega_c}{kL_{eff}z_{pk}}\delta H_x(\Delta\mu_z n + \chi_c H_z). \tag{S3}$$

To measure the *ab*-plane moment, we modulate the *c*-axis filed by $\delta H_z = 1.0$ Oe; the relative phase is also chosen to be $\theta_z = \pi/2$. The expression for the shift in dissipation is given by

$$\Delta \gamma_x \approx \frac{\omega_c}{k L_{eff} z_{pk}} (\mu_x - \chi_c H_x) \delta H_z. \quad (S4)$$

The phase-locked cantilever magnetometry technique can also be used to measure the derivative $d\mu/dH$; derivative measurements are particularly useful in studying non-hysteretic (i.e., reversible) magnetic variations; under certain conditions, derivative measurements can yield higher signal-to-noise ratio than the direct measurements of magnetic moment discussed above. In this work, we employ derivative measurements to study the high temperature behavior of the fluxoid transitions presented in Fig. 3C of the main text. For the derivative measurements, a small phase-locked modulation $\delta H_z(t)$ is applied parallel to the *c*-axis of the sample. In the presence of a non-zero static in-plane magnetic field $H_x$, the position-dependent magnetic moment produced by $\delta H_z(t)$ shifts the cantilever dissipation by

$$\Delta \gamma_x \approx \frac{\omega_c}{k L_{eff} z_{pk}} (\mu_x - \chi_c H_x) \delta H_z \quad (S5)$$

If we neglect the term $d\mu_x/dH_z$, and provided the modulation amplitude is sufficiently small, the above equation reduces to

$$\Delta \gamma_x \approx -\frac{\omega_c}{k L_{eff} z_{pk}} \left( \frac{d\mu_z}{dH_z} \right) H_x \, \delta H_z \quad (S6)$$

Here, we have expressed $\chi_c$ as $d\mu_z/dH_z$. We see from the above equation that it is possible to measure the derivative of the out-of-plane magnetic moment $d\mu_z/dH_z$ with high precision provided $H_x$ is sufficiently large.

**Supporting Text:**

## In-plane vortex penetration $H_{c1} \parallel ab$

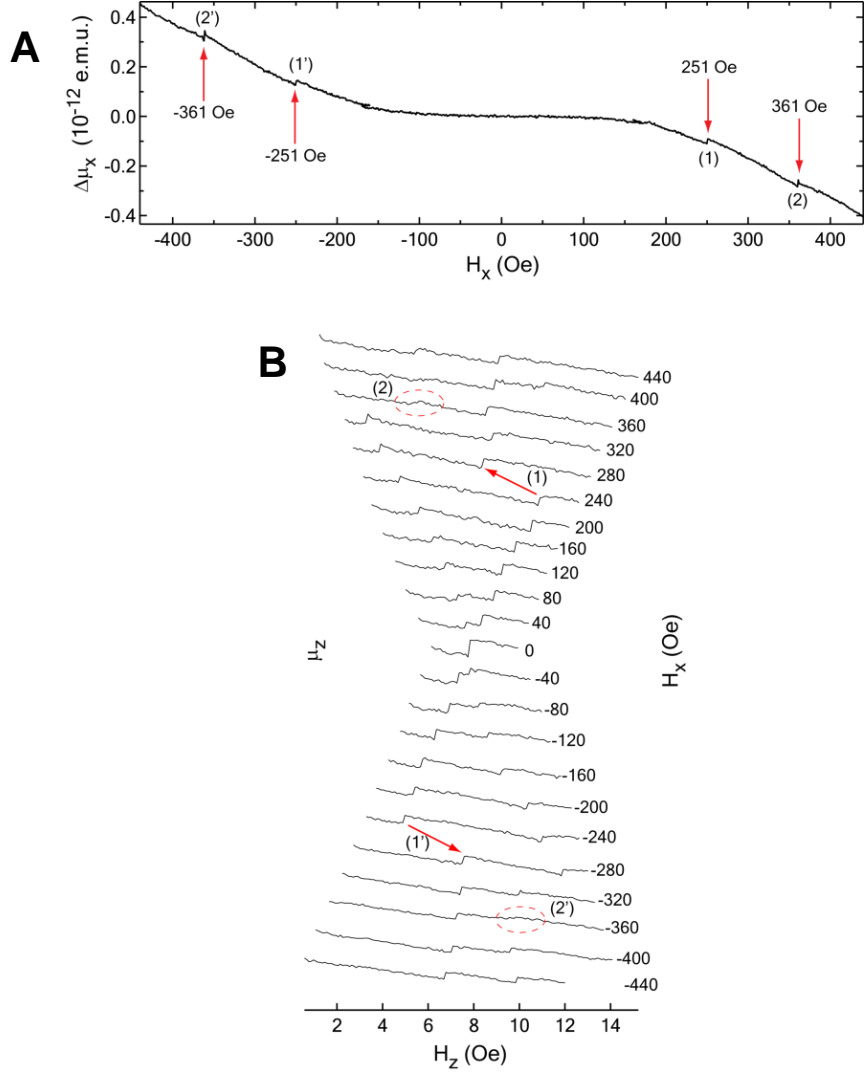

Figure S3: (**A**) In-plane magnetic moment as a function of $H_x$ obtained for the sample presented in Fig. 1 of the main text. Data was obtained for $H_z = 0$ and $T = 0.5$ K. The Meissner response has been subtracted to reveal the steps in the in-plane moment corresponding to the entry of vortices parallel to the direction of the in-plane field (red arrows). (**B**) Out-of-plane magnetic response as a function of $H_z$ and $H_x$ measured at $T = 0.5$ K. The data presented in (**A**) and (**B**) are field-cooled measurements. Accompanying the entry of in-plane vortices, we observe a shift in position (indicated by the red arrows) of the fluxoid transitions, as well as a decrease in the stability region of the HI state.

Here we present a theoretical estimate for $H_{c1} \parallel ab$ for a mesoscopic geometry taking into account the anisotropy in the superconducting response along the *c* and *ab* directions. We then compare this estimate to experimental results obtained for the sample presented in the main paper. The energetics of equilibrium vortex entry into a superconductor is affected by the geometry of the sample. In particular, if the magnetic field is applied parallel to the surface of a SC film, $H_{c1}$ is increased above the bulk $H_{c1}$ value (*S2*). This enhancement in $H_{c1}$ increases as the penetration depth increases relative to the thickness of the film. Roughly speaking, while the energy cost of the vortex core is nearly the same, in thinner films, a vortex saves a smaller amount of magnetic energy.

To obtain a theoretical estimate for $H_{c1} \parallel ab$ and $\Delta\mu_x$ (magnitude of the jump in magnetic moment associated with the entry of an in-plane vortex), we consider a superconducting rectangular box of dimension $L_x \times L_y \times h$, where $h$ is the height along the *c*-axis and the magnetic field is applied along the *x*-axis. To model the properties of SRO the magnetic response of the box is assumed to be anisotropic with $\gamma_{SC} = \lambda_c(0)/\lambda_{ab}(0) \approx 20$, where $\lambda_c(0)$ and $\lambda_{ab}(0)$ are the zero temperature *c*-axis and *ab*-plane penetration depths, respectively (*S1*). We note that, for the SRO sample in Fig. 1 of the main text, the longest dimension is approximately 1.8 μm, which is nearly a factor of two smaller than $\lambda_c(0)$; thus, we expect very little screening of $H_x$. In the limit $L_y \ll \lambda_c$ and $L_y \ll \gamma_{SC} h$, we obtain the following expressions: $H_{c1} \parallel ab = 2\gamma_{SC}\Phi_0 \ln(L_y/(\pi\xi_{ab}))/(\pi L_y^2)$ and $\Delta\mu_x = \Phi_0 L_y \Omega/(32\pi h \lambda_c^2)$, where $\xi_{ab}$ is the in-plane coherence length and $\Omega$ is the volume of the superconductor. For a box with $L_y = 1.8$ μm, $h = 0.35$ μm, $\Omega = 0.45$ μm³ and $\lambda_c = 4.3$ μm, and $\xi_{ab} = 90$ nm—corresponding to the superconducting parameters at $T = 0.4$ K, our calculations yield $H_{c1} \parallel ab \approx 150$ Oe and $\Delta\mu_x \approx 4 \times 10^{-14}$ e.m.u..

We compare this estimate to data obtained for the in-plane moment $\mu_x$ shown in Fig. S3; for details on the in-plane moment measurement, see Materials and Methods. For this sample, we find the first vortex entry occurs at $H_{c1} \parallel ab \approx 250$ Oe; the measured jump in moment is $\Delta\mu_x \approx 2.2 \times 10^{-14}$ e.m.u., both values in reasonable agreement with our estimates. We note that the observed value of $H_{c1} \parallel ab$ for this sample is much higher than the value of $H_{c1} \parallel ab \approx 8$ Oe for bulk SRO (*S3*) as expected for such mesoscopic geometries.

## Effect of rotating the direction of the in-plane magnetic field

To investigate the dependence of the half-integer (HI) state on the direction of the in-plane magnetic field, we applied the in-plane magnetic field $H_x$ along two directions rotated by 35° (see Fig. S4). Our goal was to verify whether the direction $H_x$ is important in the half-integer fraction observed in the magnetization steps accompanying fluxoid transitions as well as the dependence on the stability region of the HI state (i.e., the range of $H_z$ for which the HI state is the equilibrium state) on the magnitude of the in-plane field. We find that the half-step features persist and, furthermore, that the field-dependence of the stability region is not strongly influenced by the direction of the in-plane field.

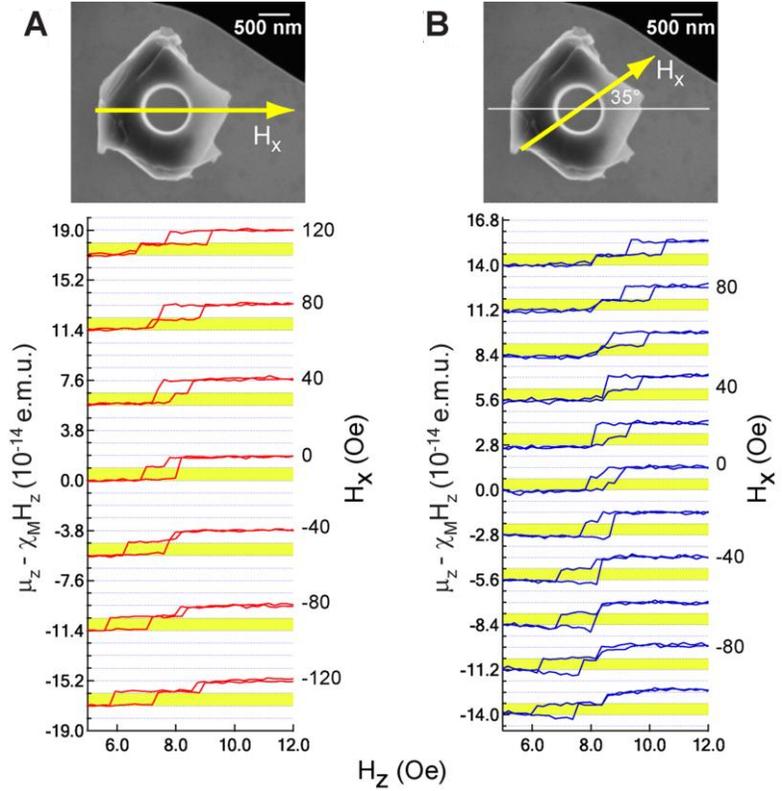

Figure S4: (**A**-**B**) In-plane field dependence of the $n=0 \rightarrow n=1$ transition measured at $T = 0.5$ K—the curves have been offset for clarity. The data were obtained by zero field cooling the sample and performing a cyclic field sweep starting from $H_z = 0$. To vary the direction of $H_x$, the cantilever was rotated in the $xy$-plane, yielding the data shown in (**B**). The data shown in (**A**) and (**B**) were obtained in separate experimental runs. To quantify the change in moment associated with the observed transitions, we obtain $\chi_M$ by fitting to the linear Messiner response between $-8\,\text{Oe} < H_z < 8\,\text{Oe}$ for the $H_x = 0$ data and then subtract $\chi_M H_z$ from the data measured at different values of $H_x$.

## Possible scenarios for the HI state

In this supplemental section we discuss three possible scenarios for the theoretical interpretation of the HI state: A kinematic spin polarization scenario in a half-quantum vortex state, a $\pi$-junction scenario, and a "wall vortex" scenario. We argue that the first scenario is more consistent with the magnitude of the observed in-plane field dependence of the HI state, whereas the last two scenarios are not consistent with our observations.

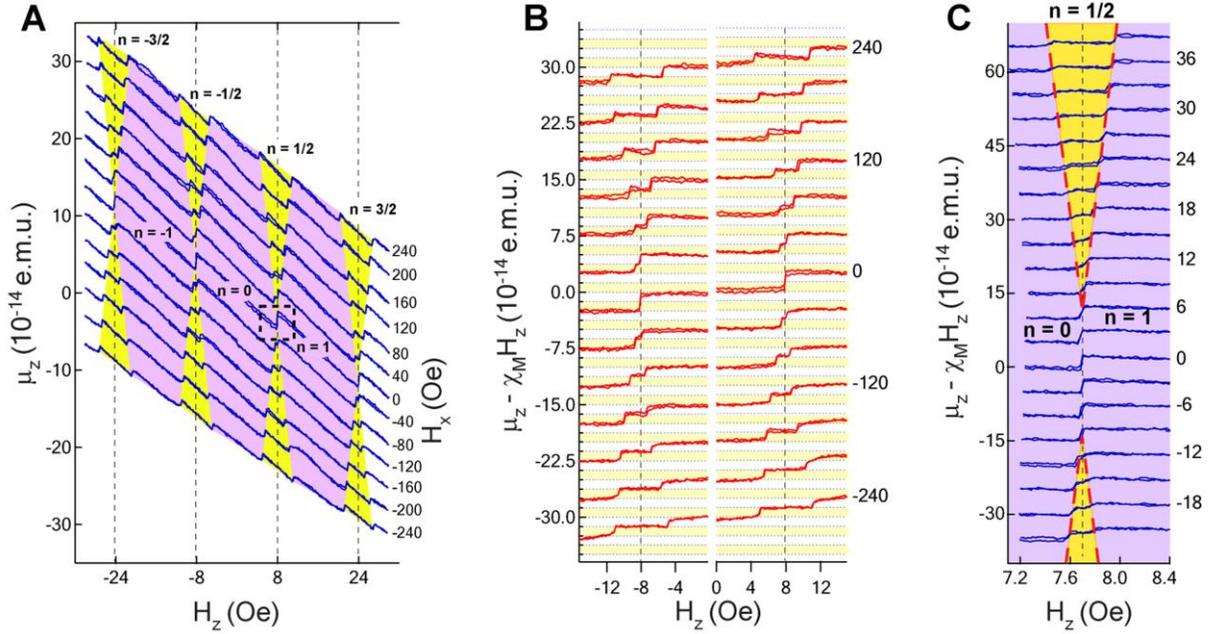

Figure S5: Plots showing zero-field-cooled data obtained at $T = 0.6$ K for the sample shown in Fig. 1 of the main text. The data shown in (**A**) represent the full magnetic response of the particle. The purple shaded regions indicate IF states, whereas the yellow shading indicates the regions in which the HI state is stable. (**B**) The data shown in (**A**) after subtraction of the linear Miessner response. (**C**) Data taken near the $n = 0 \rightarrow n = 1$ transition for the range of in-plane fields indicated by the dashed box in (**A**)—the Meissner response has been subtracted from each curve. The data shown in (**B**-**C**) have been offset for clarity by an amount proportional to $H_x$.

Based on the measurements of the SRO samples, we construct the following total Gibbs free energy (*S4*) which captures the main features of the observed magnetic response:

$$G(\boldsymbol{H}, n) = E_0(n - H_z/\Delta H_z)^2 - \bar{\chi} H_z^2/2 + E_{\text{HI}} - \boldsymbol{\mu}_{\text{HI}} \cdot \boldsymbol{H}, \tag{S7}$$

where $n$ can be integral or half-integral (i.e., HI states) (cf. Fig. S5A), and the parameters $E_0$ and $\bar{\chi}$ are expressible via measured quantities: $E_0 = \Delta\mu_z \Delta H_z/2$ and $\bar{\chi} = \chi_{\text{M}} + \Delta\mu_z/\Delta H_z$. Note that to account for the growth of the stability region of the HI states with in-plane field we have included two terms that are only nonzero in HI states: $-\boldsymbol{\mu}_{\text{HI}} \cdot \boldsymbol{H}$, where $\boldsymbol{\mu}_{\text{HI}}$ is a magnetic moment that exists only in the HI state and points in the direction of the in-plane field; and a field-independent constant contribution $E_{\text{HI}}$. From Eq. (S7), we can relate the model parameters to the growth in the stability region of the HI state as a function of $H_x$: $\delta H_z = 4\mu_{\text{HI}}(H_x - H_{x,\text{min}})/\Delta\mu_z$ (for $H_x > H_{x,\text{min}}$) and $E_{\text{HI}} = E_0/4 + \mu_{\text{HI}} H_{x,\text{min}}$. For the data shown in Fig. S5, we find $E_0 \approx 0.1$ eV, $\Delta\mu_z \approx 2 \times 10^{-14}$ e.m.u., $\mu_{\text{HI}} \approx 1 \times 10^{-16}$ e.m.u., and $E_{\text{HI}}/E_0 \approx 0.26$. We note that the value of $\mu_{\text{HI}}$ obtained is roughly 200 times smaller than the magnitude of the magnetic moment measured for an in-plane vortex ($\Delta\mu_x = 2.2 \times 10^{-14}$ e.m.u.).

**The kinematic spin polarization scenario in a half-quantum vortex state**

As previously stated, in the linear regime—provided the ESP description holds—we expect the height of the steps in the magnetic moment that occur at the transition to the HF state to be half of those that occur at IF transitions. A robust halving of the magnetization steps, which follows naturally from the theoretical framework of HQVs, is what we observe throughout the range of fields and temperatures studied. Perhaps the most intriguing observation is of an in-plane magnetic moment $\boldsymbol{\mu}_{\text{HI}}$ in the HI state. While the origin of this moment is, as yet, unknown, recent work by Vakaryuk and Leggett (*S5*) finds that a kinematic spin polarization $\boldsymbol{\mu}_s^{\text{kin}}$ can develop in the HQV state, as a result of the velocity mismatch between the $|\uparrow\uparrow\rangle$ and $|\downarrow\downarrow\rangle$ superfluid components. While, theoretically, the magnitude of $\boldsymbol{\mu}_s^{\text{kin}}$ depends on the distribution of both charge and spin currents, we estimate it to be of the same order of magnitude as the experimental value of $\mu_{\text{HI}} \approx 1 \times 10^{-16}$ e.m.u. (see the next section of the SOM for more details).

To complete the interpretation of the experimental data, described by Eq. (*S7*), within the framework of HQVs, we consider two additional factors that are primarily responsible for the free-energy difference between the HF and IF states captured by the term $E_{HI}$ in Eq. (S7): one arises from the presence of spin currents; the other from "spin-orbit" interactions (*S6*), which depend on the orientation of $\boldsymbol{d}$ with respect to the internal Cooper-pair angular momentum for the full and half quantum vortex states. The contribution from the spin currents is determined by their spatial distribution, and is proportional to the spin superfluid density $\rho_{sp}$ (*S7-S8*). To evaluate the contribution of spin-orbit interactions, both the orientation of $\boldsymbol{d}$ and the spin-orbit coupling strength are needed—neither have been experimentally determined. Theoretically, there are several models for the order parameter of SRO which predict distinct orientations for $\boldsymbol{d}$ (see e.g. (*S1, S9-S10*)). Our interpretation of $\boldsymbol{\mu}_{HI}$ as being caused by a spin polarization implies that the ESP axis should be in the direction of the in-plane field for the range of fields studied.

**The $\pi$-junction scenario**

Under certain circumstances, there can exist crystal grain boundary Josephson junctions within the sample that can shift the total phase winding by $\pi$ (*S11*). However, our samples are fabricated from high quality single crystals, and thus it is unlikely that they contain the grain boundaries and orientations necessary to realize a $\pi$-junction. Moreover, even assuming that the presence of a $\pi$-junction, such as in (*S11*), its impact would be to shift the values of the out-of-plane magnetic field at which the fluxoid transitions occur by $\Delta H_z/2$. We do not find evidence of such a shift in our data. Furthermore, the $\pi$-junction scenario would produce a jump in magnetic moment identical to that for a conventional full-quantum vortex. It would not lead to the half-height jumps of the magnetic moment. Therefore, it seems implausible that such a scenario is relevant to our observations.

**The wall vortex scenario**

We refer to a wall vortex (WV) as any state where a vortex penetrates through the volume of the sample. In general, a transition between a WV state and the $n = 0$ integer fluxoid state will correspond to a change in the particle's magnetic moment $\Delta\boldsymbol{\mu}_{WV}$. Thus, to interpret the HI state as a WV state, $\Delta\boldsymbol{\mu}_{WV}$ would need to be consistent with the observed value of $(\Delta\mu_z/2)\hat{z} + \boldsymbol{\mu}_{HI}$. For the experimental particles where the HI state is observed, the local superconducting properties are not known in detail. Therefore it is difficult to constrain the possible forms of WV states. However, we can make the following general observations: (i) For the ob-served in-plane vortices corresponding to fields $H_x \geq 250$ Oe (Fig. S3), the magnitude of the in-plane component of $\Delta\boldsymbol{\mu}_{WV}$ is approximately 200 times larger than $\mu_{HI}$. Thus, the HI state is not a simple generalization of the observed in-plane WV state. (ii) The z-axis component of $\Delta\boldsymbol{\mu}_{WV}$ can be any fraction of $\Delta\mu_z$ and is not generically $\Delta\mu_z/2$. (iii) In general, we would expect the location/orientation of wall vortices to vary with the magnitude and direction of the applied field. Hence, we would expect multiple fractional steps in the magnetic moment; these steps correspond to transitions involving various WV states. (iv) Given the geometric asymmetry of our samples, we would also expect that the component of $\Delta\boldsymbol{\mu}_{WV}$ along the in-plane field to vary with the direction of the in-plane field. Consequently, the stability region of the WV state should be affected by the direction of the in-plane field. However, over the range of fields studied for the SRO particle shown in Fig. 1 of the main text, we find that the stability region of the HI feature is not influenced by the direction of the in-plane field (Fig. S4). Given these considerations we conclude that to formulate a wall vortex scenario consistent with the observed properties of the HI state would require a fine tuned set of assumptions, and is thus unlikely.

## Effect of kinematic spin polarization on the stability of HQV state

As pointed out in (*S5*), the velocity mismatch between $|\uparrow\uparrow\rangle$ and $|\downarrow\downarrow\rangle$ spin components of an ESP superfluid in the HQV state gives rise to an effective Zeeman field $B_{eff}$ which, in thermal equilibrium, produces a kinematic spin polarization in addition to that caused by the Zeeman coupling to the external field. Such kinematic spin polarization is absent in the full vortex state where the velocities of $|\uparrow\uparrow\rangle$ and $|\downarrow\downarrow\rangle$ components are the same. The goal of this section is to argue that the coupling between the kinematic spin polarization and the external field can account for the experimentally observed growth of the stability region of half-integer steps. The detailed calculation of the effect of the kinematic spin polarization on the stability of HQVs will be given elsewhere.

Generalizing results of ref. (*S5*) for an ESP superconductor of arbitrary geometry the effective Zeeman field in the HQV state becomes a local vector which points along the ESP axis and its magnitude is given by

$$B_{eff}(\boldsymbol{r}, \boldsymbol{H}) = -\frac{m^*}{g_S \mu_B}\left(1 + \frac{F_1}{3} + \frac{Z_1}{12}\right) \boldsymbol{v}_s(\boldsymbol{r}, \boldsymbol{H}) \cdot \boldsymbol{v}_{sp}(\boldsymbol{r}) \qquad (S8)$$

where $m^*$ is the Fermi-liquid mass of the charge carriers and $g_S$ is their gyromagnetic ratio. Fermi-liquid parameters $F_1$ and $Z_1$ describe renormalization of charge and spin currents respectively; the notation for Fermi liquid parameters is the same as in (*S7*). In the expression (S8) $\boldsymbol{v}_s$ and $\boldsymbol{v}_{sp}$ are the local charge and spin superfluid velocities respectively; the superfluid velocity $v_s$ which describes motion of charges in the system couples to the applied field $\boldsymbol{H}$.

In our experimental setup the applied field $\boldsymbol{H}$ has both in-plane ($H_x$) and $c$-axis ($H_z$) components. However, because of the large superconducting and shape anisotropies of the sample, the measured magnetic response, and hence the current distribution, is dominated by the $c$-axis field. Assuming that the ESP axis lies in the *ab* plane, the Gibbs potential of full and half-quantum vortices can be written as

$$G^{FQV}(\boldsymbol{H}) = G_0^{FQV}(H_z) - \frac{1}{2}\chi_S \int \frac{d^3\boldsymbol{r}}{\Omega} H_x^2 \qquad (S9)$$

$$G^{HQV}(\boldsymbol{H}) = G_0^{HQV}(H_z) - \frac{1}{2}\chi_S \int \frac{d^3\boldsymbol{r}}{\Omega} \{H_x - B_{eff}(\boldsymbol{r}, H_z)\}^2$$

where only spin-polarization parts of the energies have been explicitly written. Here, $\chi_S$ and $\Omega$ are the spin susceptibility and sample volume, respectively. The region of the thermodynamic stability of the HQV as a function of the applied field $\boldsymbol{H}$ is determined by equating the above Gibbs potentials which leads to the following equation

$$\chi_S H_x \int \frac{d^3\boldsymbol{r}}{\Omega} B_{eff}(\boldsymbol{r}, H_z) = \delta G(H_z) \qquad (S10)$$

It should be emphasized that the dependence on the in-plane field $H_x$ in the above equation enters only linearly through the l.h.s., as explicitly indicated, while all other energy contributions are gathered on the r.h.s. and are denoted as $\delta G(H_z)$.

To analyze the dependence of the $c$-axis field stability region of the HQV on the in-plane field $H_x$ as implicitly determined by Eqn. (S10) we note that the kinematics of charge and spin currents favors the stability region to be located around the field corresponding to the transition between full vortex states. Denoting this field as $H_z^{(0)}$ and expanding both $B_{eff}(\boldsymbol{r}, H_z)$ and $\delta G(H_z)$ in its neighborhood we obtain that the $c$-axis field stability region of HQV depends linearly on the in-plane field with the slope given by

$$\left.\frac{dH_z}{dH_\text{x}}\right|_{FQV \to HQV} = \frac{\chi_s B_{eff}}{\Delta M}\zeta \qquad (S11)$$

where $\Delta M$ is the magnetization jump between full and half-quantum vortex states as derived from $\delta G(H_z)$; parameters $\zeta$ and $B_{eff}$ are defined below. Denoting the characteristic size of the sample as $R$, the dimensionless parameter $\zeta$ is given by

$$\zeta \equiv \left[\frac{\hbar}{2m^*R}\right]^{-2} \int \frac{d^3\mathbf{r}}{\Omega} \mathbf{v}_s\left(\mathbf{r}, H_z^{(0)}\right) \cdot \mathbf{v}_{sp}(\mathbf{r}) \qquad (S12)$$

and characterizes the overlap of the charge and spin currents. Although the exact distributions of both currents for our sample are unknown, the important observation is that for a mesoscopic sample with dimensions comparable to the London penetration depth the overlap integral which determines $\zeta$ need not be small.

The other parameter in Eqn. (S12) is a constant field $B_{eff}$ defined as

$$B_{eff} \equiv \frac{1}{2g_S}\left[1 + \frac{Z_1/12}{1 + F_1/3}\right]\frac{\phi_0}{\pi R^2} \qquad (S13)$$

Given the free-electron value of the gyromagnetic ratio $g_S = 2$, the size of our sample and the assumption that Fermi liquid parameter $Z_1$ is small (as it is for $^3$He) $B_{eff}$ is estimated to be of order 5 G. The crucial observation is that the magnitude of $B_{eff}$ depends substantially on the size of the sample.

It is convenient to transform the overlap integral in $\zeta$ to a form which involves a surface integral. Given that (i) the spin superfluid velocity $\mathbf{v}_{sp}$ is described by a potential flow, and (ii) superfluid velocity $\mathbf{v}_s$ is proportional to the superconducting current density, one obtains the following alternative form of $\zeta$:

$$\zeta = \frac{m^*R}{\hbar}\frac{\lambda_L^2}{\phi_0}\int \frac{R\,d^2\mathbf{S}}{\pi\Omega}\cdot(\mathbf{H}\times\mathbf{v}_{sp}) \qquad (S14)$$

where $\mathbf{H}$ is the total magnetic field on the surface of superconductor. However, the condition of the potentiality of the spin superflow leads to the conclusion that a part of $\mathbf{H}$ which is constant and uniform (e.g. the applied field) does not contribute to the value of $\zeta$ so that the field $\mathbf{H}$ in Eqn. (S14) can be treated as the one generated by the superconductor only.

To obtain an estimate of $\zeta$ we consider a doughnut-shaped superconducting sample with the mean and cross-section radii denoted as $R$ and $r$, respectively. For this geometry, the magnetic field on the surface is given by $2M/\pi R^2 r$, where $M$ is the magnetic moment of the doughnut, and, provided the rotation of the **d** is uniform, the average value of the spin current in the HQV state is given by $\hbar/4m^*R$. This leads to the following expression for $\zeta$:

$$\zeta \approx 8\pi^2 \frac{\lambda_L^2 M}{\phi_0 \Omega} \tag{S15}$$

The magnetic moment $M$, which is determined by the applied field $H_z$, depends also on the geometry of the sample and on the value of the penetration depth. For a small sample, the latter dependence can be factored out and $M$ can be approximated by $M \approx H_z \Omega R^2/2\pi\lambda_L^2$ which gives $\zeta \sim 1$ for the values of the applied fields typical for our experiment.

We can now estimate the slope of the transition line between the full and half quantum vortex states as given by Eqn. (S11) and also the magnitude of the kinematic spin polarization which follows from it. Using the estimates for $B_{eff}$ and $\zeta$ obtained above, $\Delta M \approx 10^{-14}$ e.m.u and $\chi_S \approx 2 \times 10^{-17}$ cm$^3$, which is obtained from the molar spin susceptibility of SRO (i.e. $10^{-3}$ e.m.u./mol (*S12*)) and the volume of the sample, we find $\left.\frac{dH_z}{dH_x}\right|_{FQV \to HQV} \approx 0.01$ and $\mu_S^{kin} \approx 10^{-16}$ e.m.u., in agreement with the experimentally observed value.

## Edge currents from chiral domains

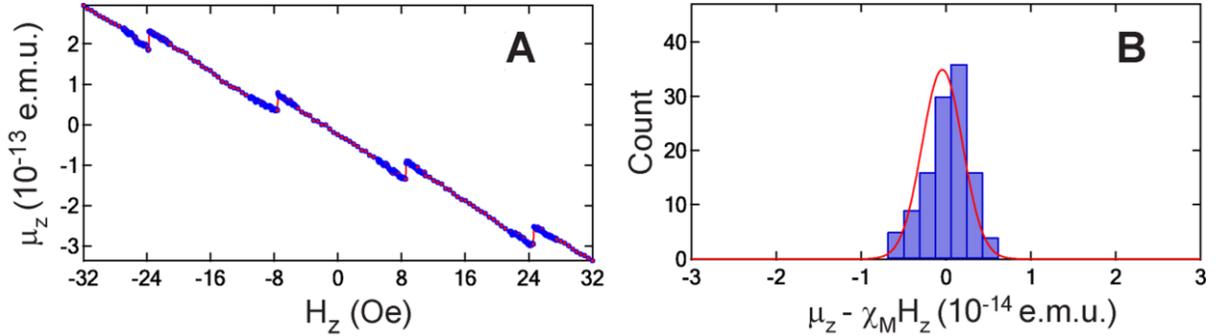

Figure S6: (**A**) Field-cooled data obtained at $T = 0.4$ K for $H_x = 0$. To quantify the fluctuations in the magnetic moment $\mu_z$, we subtract the Meissner response and histogram the difference $\mu_z - \chi_M H_z$. The histogram shown in (**B**) represent the fluctuations for field-cooled data between $-8\ Oe < H_z < 8$ Oe; the standard deviation fluctuations is $\sigma_z = 2.5 \times 10^{-15}$ e.m.u..

To compare the in-plane orientation of the ESP axis that our interpretation of the data implies with the orientation proposed in the literature (*S1*), we note that of the five possible unitary states that lack symmetry required nodes, only the chiral states, which have the schematic form $\hat{\boldsymbol{d}} = \hat{z}(p_x \pm i\, p_y)$, have their ESP axis in-plane. Such chiral states have been predicted to carry spontaneous edge-currents of magnitude $5.6 \times 10^{-6}$ A per layer (*S13-S14*). If the sample has a single chiral domain then, together, the chiral currents and the Meissner screening currents, would produce an additional magnetic moment along the c-axis, of order $\pm 10^{-12}$ e.m.u. However, to within the noise resolution of the experiment, the field-cooled magnetization curve of Fig. S6A shows no signature of a zero-field moment, which limits its observed value to less than $2.5 \times 10^{-15}$ e.m.u. Several effects may account for this discrepancy: (i) a reduction of the edge currents (see e.g. (*S10, S15*)), (ii) the presence of multiple domains (*S16*), or (iii) insufficient thermalization of the particle during the field-cooling procedure. To resolve this discrepancy, a more systematic study is needed. We emphasize, however, that the existence of HQVs requires only ESP. Thus, even if we accept the basic idea that the observation of an HI state would infer the existence of an (underlying) HF state, further reasoning is required to pin down the precise form of the superconducting state.

# SRO Sample #2

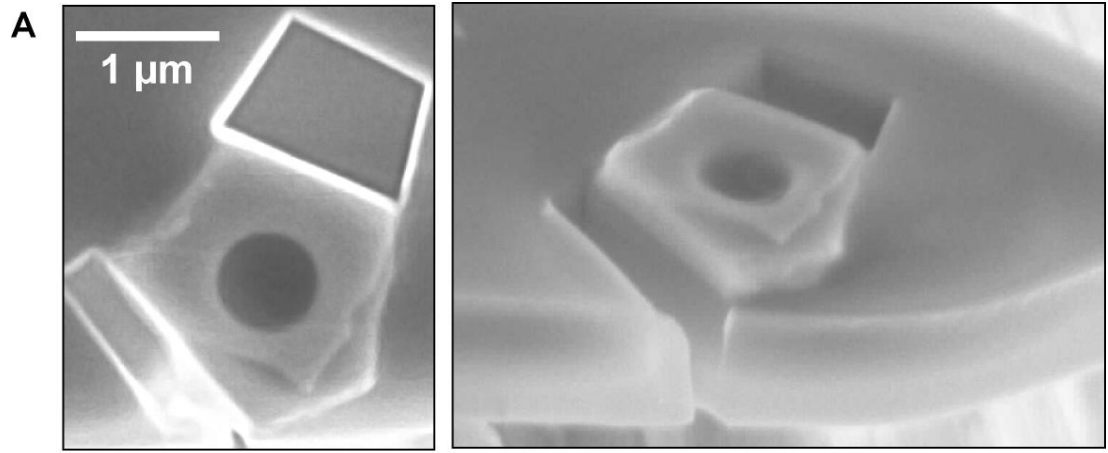

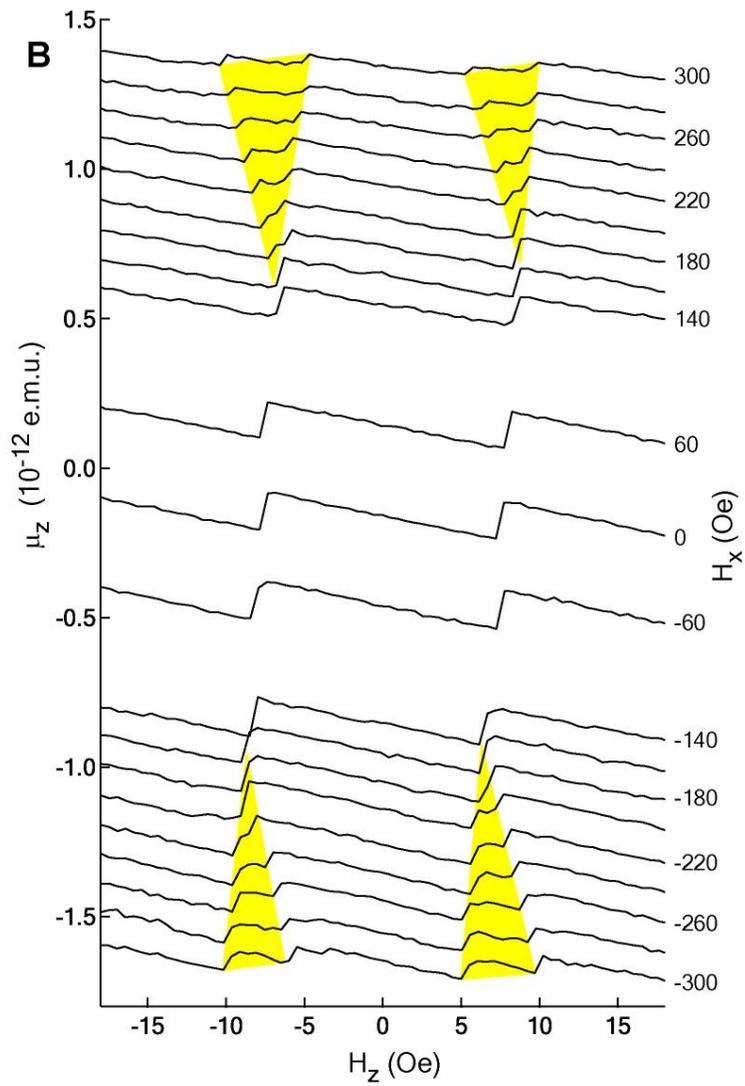

# SRO Sample #2

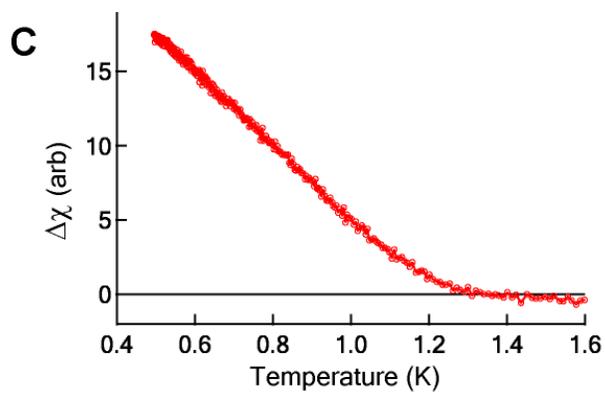

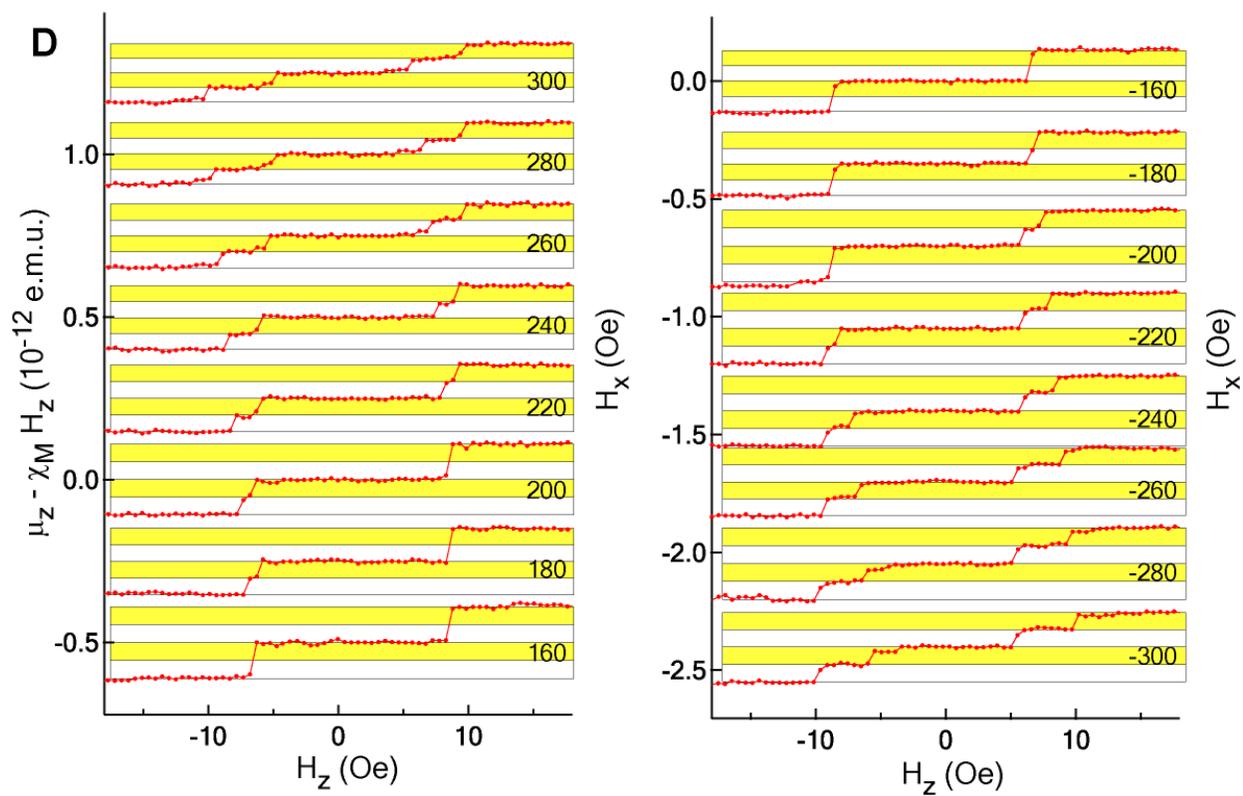

# SRO Sample #2

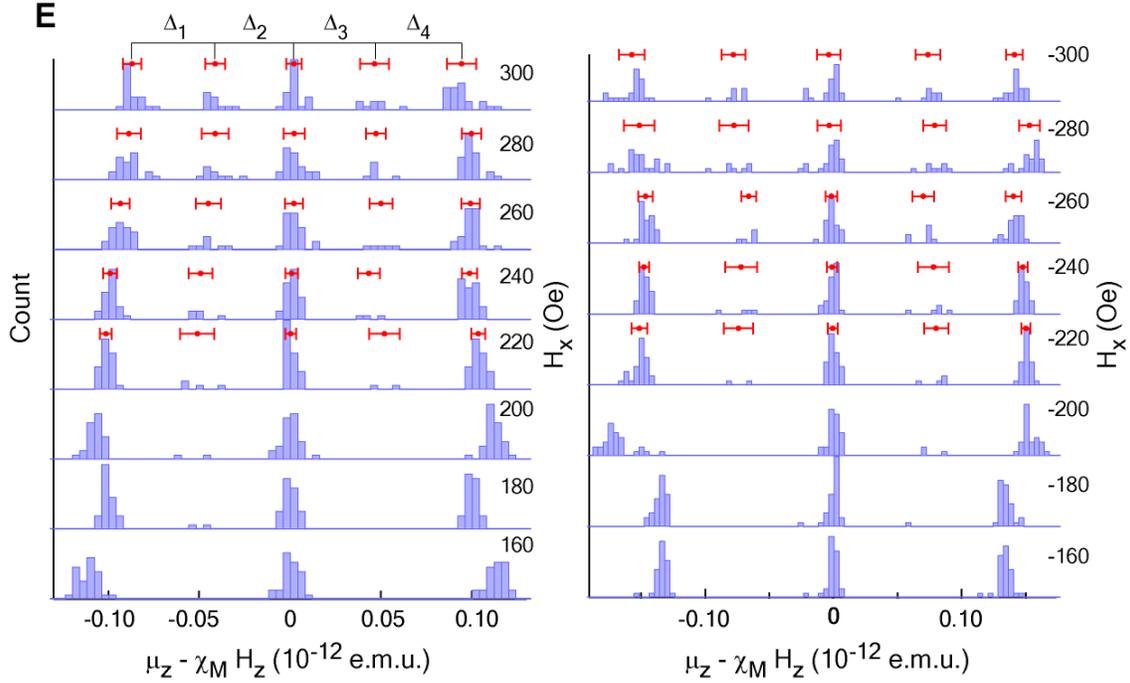

Figure S7: (**A**) SEM of a 0.35-µm thick annular SRO particle with a 0.75-µm diameter hole. The SRO particle is shown glued to a silicon cantilever, the notches in the cantilever are a result of the focused ion beam shaping of the particle. To obtain the equilibrium magnetic response, the particle is field cooled at each value of the applied field. (**B**) Equilibrium data obtained at $T = 0.8\ K$; each curve has been offset for clarity. (**C**) The susceptibility of the particle as a function of temperature. (**D**) Meissner-subtracted data. (**E**) Histograms of the Meissner-subtracted data.

| $H_x$ (Oe) | $\Delta_1/(\Delta_1 + \Delta_2)$ | $\Delta_3/(\Delta_3 + \Delta_4)$ |
|---|---|---|
| 300 | $0.51 \pm 0.06$ | $0.48 \pm 0.08$ |
| 280 | $0.52 \pm 0.07$ | $0.46 \pm 0.06$ |
| 260 | $0.51 \pm 0.06$ | $0.49 \pm 0.06$ |
| 240 | $0.50 \pm 0.05$ | $0.43 \pm 0.05$ |
| 220 | $0.49 \pm 0.07$ | $0.50 \pm 0.06$ |
| -220 | $0.51 \pm 0.06$ | $0.53 \pm 0.05$ |
| -240 | $0.52 \pm 0.06$ | $0.53 \pm 0.06$ |
| -260 | $0.56 \pm 0.04$ | $0.50 \pm 0.05$ |
| -280 | $0.50 \pm 0.07$ | $0.53 \pm 0.06$ |
| -300 | $0.52 \pm 0.06$ | $0.53 \pm 0.06$ |
|  | $\langle\Delta_1/(\Delta_1 + \Delta_2)\rangle$ | $\langle\Delta_3/(\Delta_3 + \Delta_4)\rangle$ |
|  | $0.51 \pm 0.02$ | $0.50 \pm 0.02$ |

Table S1: The calculated fractional step heights for the data shown in Fig. S7E. The average value of a given fractional step is indicated by the quantity $\langle...\rangle$.

**SRO Sample #3**

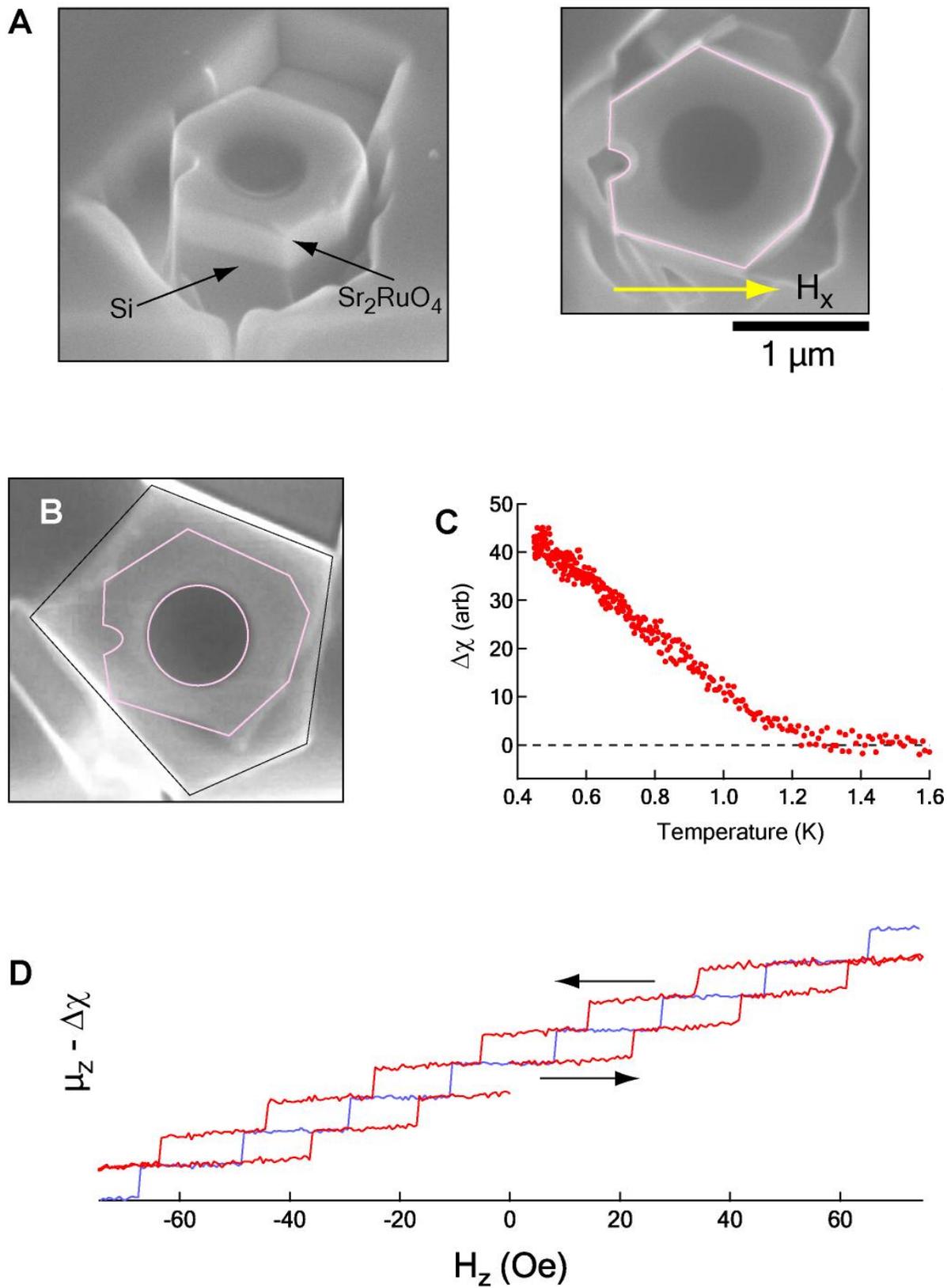

**SRO Sample #3**

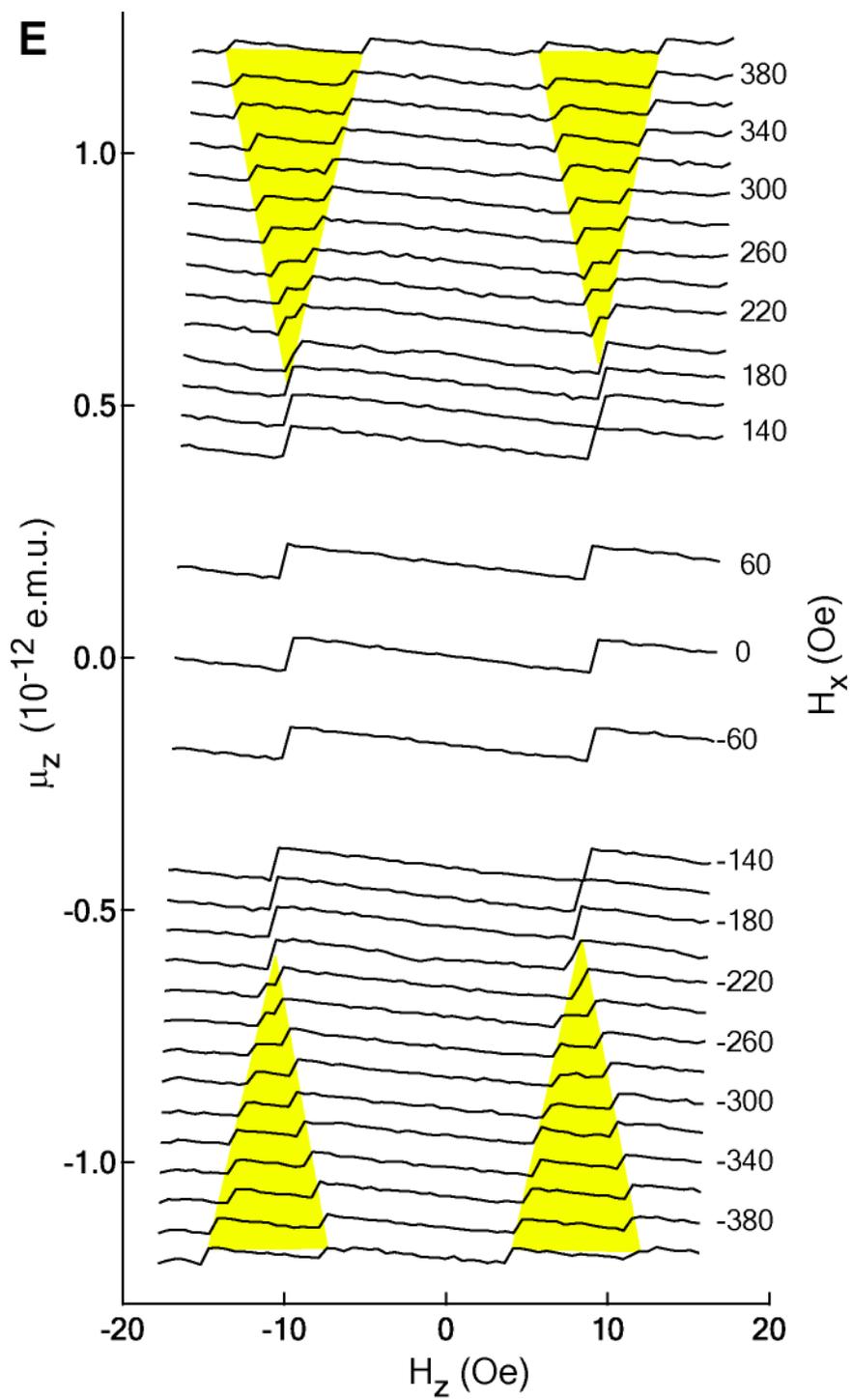

# SRO Sample #3

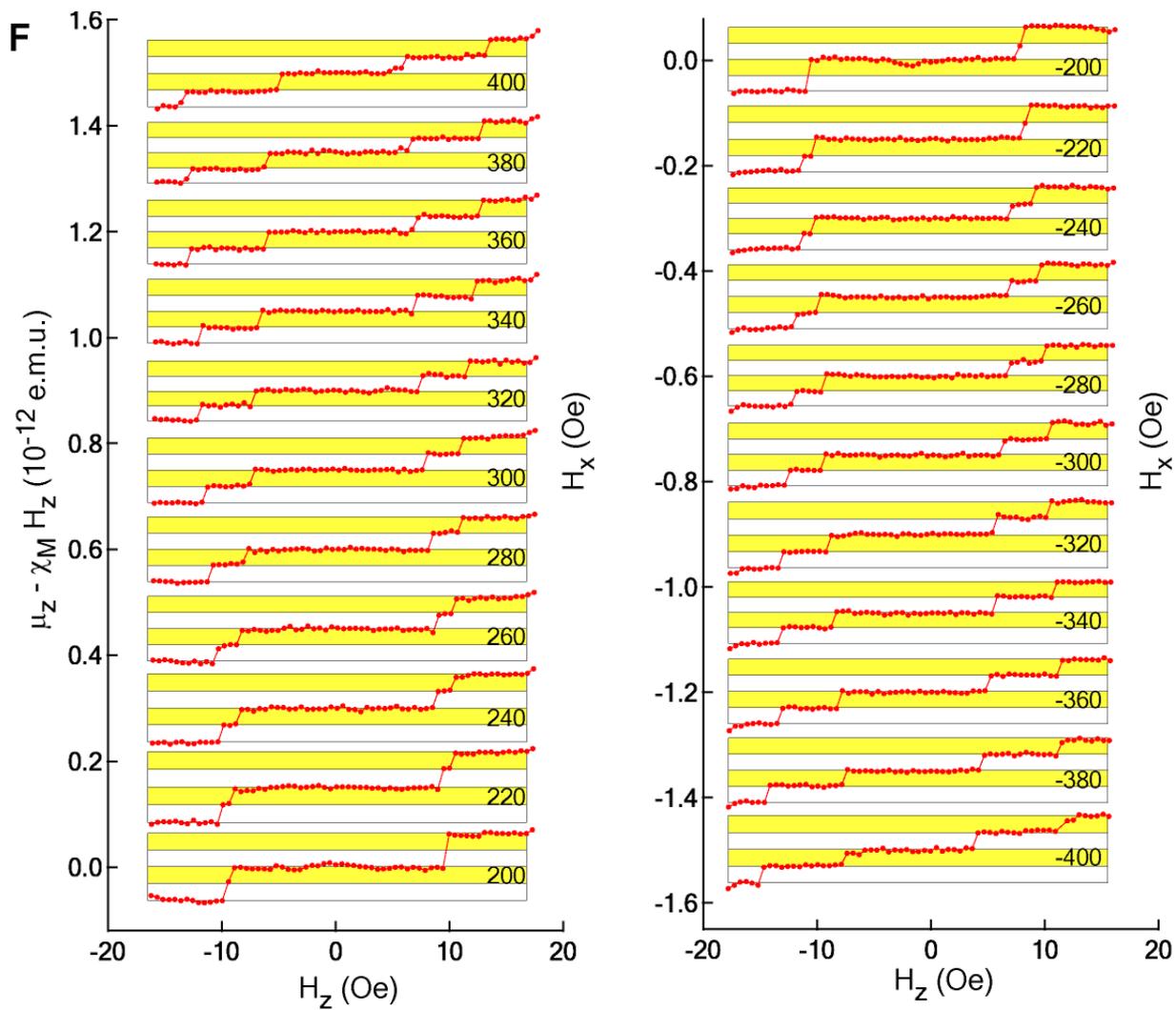

# SRO Sample #3

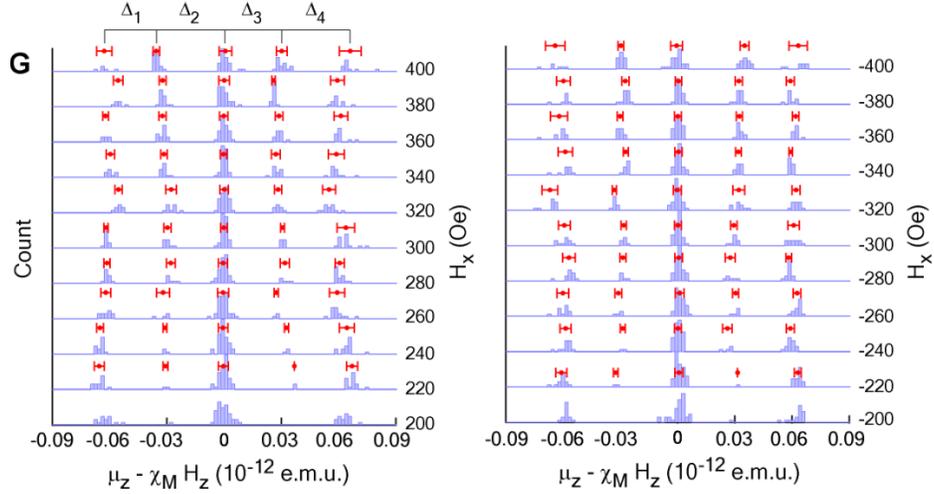

Figure S8: (**A**) SEM of the sample shown in Fig. S13 after being reshaped by the focused ion beam. (**B**) Comparison before-and-after reshaping the particle. The black outline indicates the boundary of the sample as it appears in Fig. S7A; the purple outline indicates the boundary after reshaping; the overall sample volume has been reduced to 44% of its initial volume after the reshaping. (**C**) The susceptibility of the particle after reshaping as a function of temperature. (**D**) The Meissner-subtracted equilibrium (blue curve) and zero-field-cooled data (red curve) obtained for $H_x = 0$ at $T = 0.5\,K$. We note, although the sample dimensions have been reduced considerably, the transitions are still highly hysteretic, indicating that the ring does not possess a weak link. (**E**) Equilibrium data obtained at $T = 0.5\,K$; each curve has been offset for clarity. (**F**) The Meissner-subtracted data. (**G**) Histograms of the Meissner-subtracted data.

| $H_x$ (Oe) | $\Delta_1/(\Delta_1 + \Delta_2)$ | $\Delta_3/(\Delta_3 + \Delta_4)$ |
|---|---|---|
| 400 | $0.43 \pm 0.05$ | $0.45 \pm 0.06$ |
| 380 | $0.42 \pm 0.04$ | $0.44 \pm 0.04$ |
| 360 | $0.48 \pm 0.03$ | $0.47 \pm 0.04$ |
| 340 | $0.47 \pm 0.03$ | $0.46 \pm 0.05$ |
| 320 | $0.50 \pm 0.04$ | $0.51 \pm 0.04$ |
| 300 | $0.52 \pm 0.03$ | $0.48 \pm 0.04$ |
| 280 | $0.55 \pm 0.04$ | $0.53 \pm 0.04$ |
| 260 | $0.49 \pm 0.05$ | $0.47 \pm 0.04$ |
| 240 | $0.53 \pm 0.03$ | $0.51 \pm 0.04$ |
| 220 | $0.53 \pm 0.03$ | $0.55 \pm 0.03$ |
| -220 | $0.46 \pm 0.03$ | $0.49 \pm 0.02$ |
| -240 | $0.51 \pm 0.03$ | $0.44 \pm 0.04$ |
| -260 | $0.48 \pm 0.04$ | $0.48 \pm 0.03$ |
| -280 | $0.49 \pm 0.04$ | $0.47 \pm 0.04$ |
| -300 | $0.52 \pm 0.03$ | $0.48 \pm 0.04$ |
| -320 | $0.51 \pm 0.04$ | $0.52 \pm 0.04$ |
| -340 | $0.54 \pm 0.04$ | $0.53 \pm 0.02$ |
| -360 | $0.52 \pm 0.04$ | $0.52 \pm 0.03$ |
| -380 | $0.54 \pm 0.04$ | $0.54 \pm 0.03$ |
| -400 | $0.54 \pm 0.05$ | $0.56 \pm 0.05$ |
|  | $\langle \Delta_1/(\Delta_1 + \Delta_2) \rangle$ | $\langle \Delta_3/(\Delta_3 + \Delta_4) \rangle$ |
|  | $0.50 \pm 0.01$ | $0.50 \pm 0.01$ |

Table S2: The calculated fractional step heights for the data shown in Fig. S8G. The average value of a given fractional step is indicated by the quantity $\langle ... \rangle$.

# SRO Sample #4

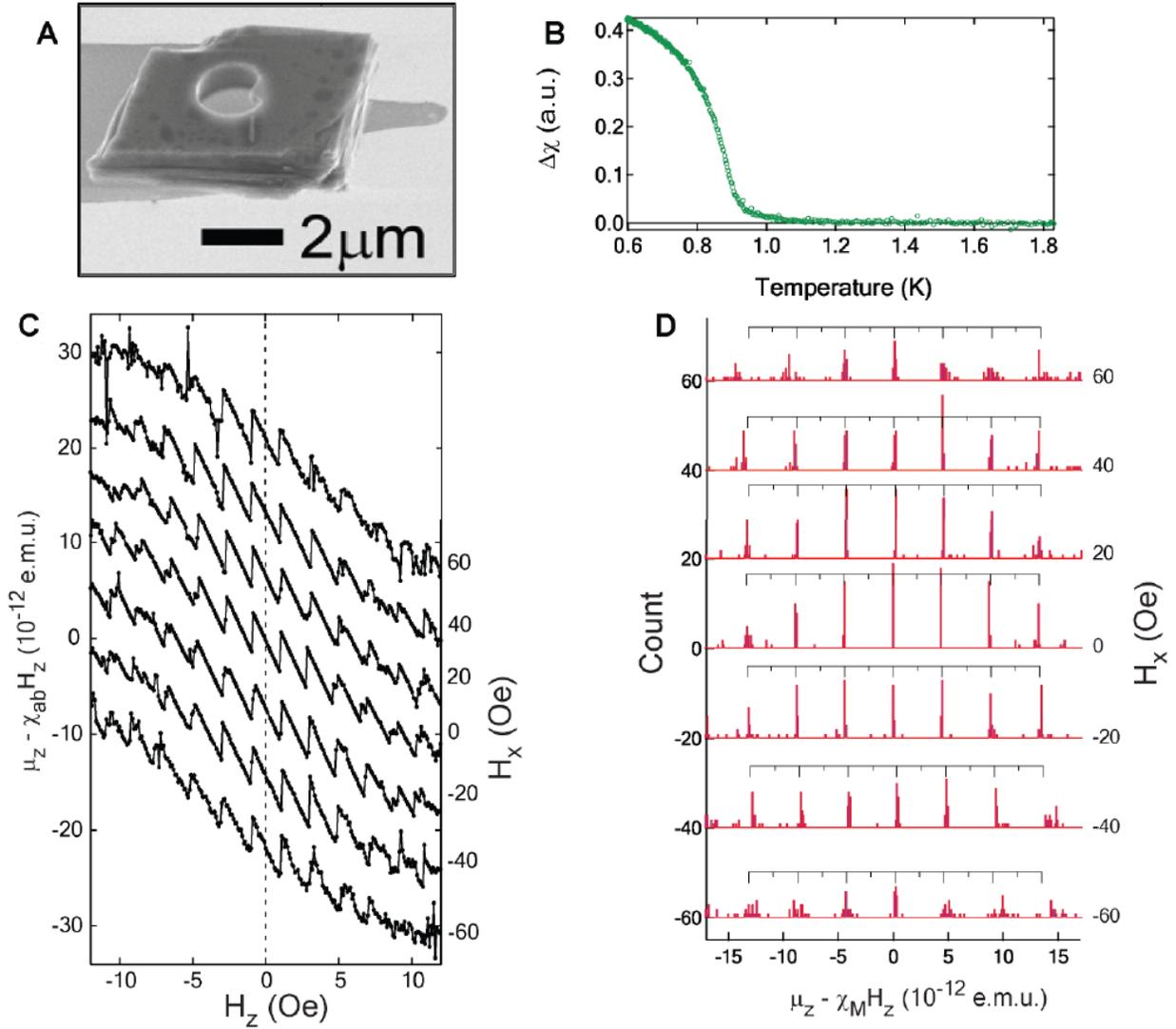

Figure S9: (**A**) SEM of a $6\ \mu m \times 4\ \mu m \times 1.6\ \mu m$ (thick) annular SRO particle with a $1.7\ \mu m$ diameter hole; the particle is shown attached to the tip of a silicon cantilever. (**B**) The susceptibility of the particle as a function of temperature. (**C**) Field-cooled data obtained at $T = 0.5$ K for values of the in-plane magnetic field between -60 Oe to 60 Oe. Each curve has been offset for clarity. (**D**) Histograms of the Meissner-subtracted data. The histograms represent data for the $n = -3$ to $n = 3$ fluxoid transitions. As a guide to the eye, we have placed a scale bar above each histogram, where the major ticks represent the values of the magnetic moment corresponding to the integer fluxoid states of the hole.

## NbSe₂ sample

For comparison, we have studied the magnetic response of an annular conventional superconductor of similar size to the SRO particle presented in the main text. We chose the s-wave superconductor NbSe₂ because, like SRO, it is a layered compound that exhibits an anisotropic superconducting response; the effective mass anisotropy in NbSe₂ is $(\lambda_c/\lambda_{ab}) \approx 4$. The c-axis penetration depth $\lambda_{ab} = 0.15\ \mu m$ is comparable to that of SRO.

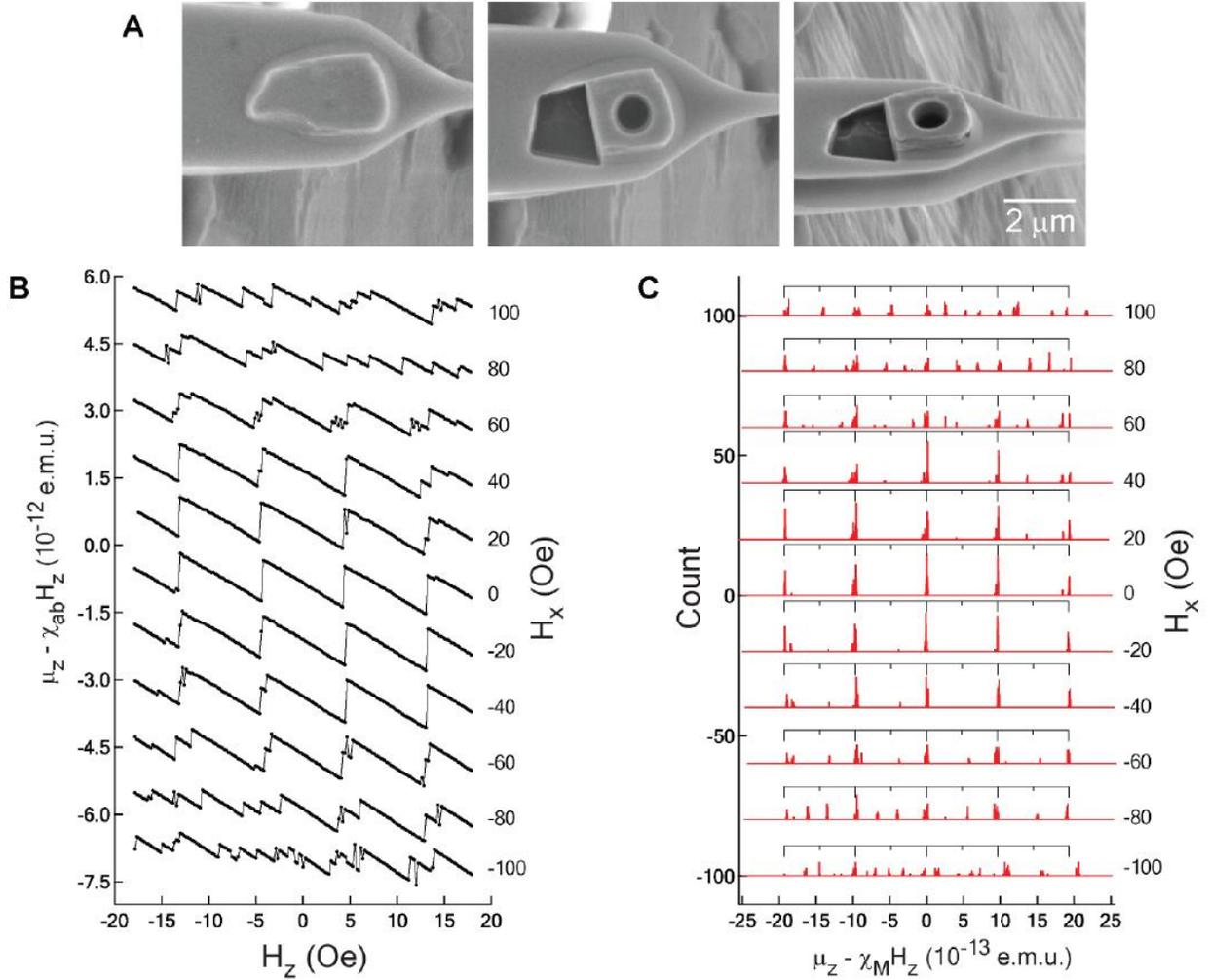

Figure S10: (**A**) SEM of a $2\ \mu m \times 2\ \mu m \times 0.3\ \mu m$ (thick) annular NbSe₂ particle with a $1.0\ \mu m$ diameter hole is shown in central and right hand panels; the image on the left shows the particle after being glued to the tip of a silicon cantilever prior to being shaped by the focused ion beam. The critical temperature of the annular particle was determined to be $T_c = 7.2$ K. (**B**) Field-cooled data obtained at $T = 6.0$ K for values of the in-plane magnetic field between -100 Oe to 100 Oe. Each curve has been offset for clarity. (**C**) Histograms of the Meissner-subtracted data;

histograms have been offset for clarity. As a guide to the eye, we have placed a scale bar above each histogram, where the major ticks represent the values of the magnetic moment corresponding to the integer fluxoid states of the hole.

**Supporting References and Notes**